\documentclass[aps,showpacs,amsmath,amssymb]{revtex4}
\usepackage{mathptmx}
\usepackage{amsmath}
\usepackage{amssymb}
\usepackage{graphicx}
\usepackage{euscript}

\newlength{\pictwidth}
\newcommand{\aver}[1]{{\left< #1 \right>}}
\normalsize
\begin{document}
\title{Ray chaos and ray clustering in an ocean waveguide} 

\author{D.V. Makarov}\email{makarov@poi.dvo.ru}
\author{M.Yu. Uleysky} 
\author{S.V. Prants}\email{prants@poi.dvo.ru}
\affiliation{V.I.Il'ichev Pacific Oceanological Institute \\
of the Russian 
Academy of Sciences, 690041 Vladivostok, Russia}

\begin{abstract}
We consider ray propagation in a waveguide with a designed
sound-speed profile perturbed by a range-dependent perturbation
caused by internal waves in deep ocean environments. The Hamiltonian
formalism in terms of the action and angle variables is applied
to study nonlinear ray dynamics with two sound-channel models
and three perturbation models:
a single-mode perturbation, a random-like sound-speed fluctuations,
and a mixed perturbation. In the integrable limit without any perturbation,
we derive analytical expressions for ray arrival times and timefronts
at a given range, the main measurable characteristics in field
experiments in the ocean. In the presence
of a single-mode perturbation, ray chaos is shown to arise as a result of 
overlapping nonlinear ray-medium resonances. Poincar\'e maps, plots of variations
of the action per a ray cycle length, and plots with rays escaping
the channel reveal inhomogeneous structure of the underlying phase space
with remarkable zones of stability where stable coherent ray clusters 
may be formed.
We demonstrate the possibility of determining the wavelength of the 
perturbation mode from the arrival time distribution under conditions
of ray chaos. It is surprising that coherent ray clusters, consisting
of fans of rays which propagate over long ranges with close dynamical
characteristics, can survive under a random-like multiplicative perturbation
modelling sound-speed fluctuations caused by a wide spectrum of internal waves.
\end{abstract}
\pacs{05.45.Ac; 05.40.Ca; 43.30.+m; 92.10.Vz}
\maketitle

\section{\label{secIntro}Introduction}

Low-frequency acoustic signals may propagate in the deep ocean
to long ranges (up to a few thousands kilometers) due to existence
of the underwater sound channel which acts as a waveguide confining
the sound waves within a restricted water volume and preventing
their interaction with the lossy ocean bottom \cite{Brekh}. In the ray approximation,
the underwater sound propagation can be modelled by a Hamiltonian
system representing a nonlinear oscillator driven by a weak nonstationary
external perturbation. A range-independent background sound speed profile
plays the role of an unperturbed potential on which a range-dependent
perturbation of the sound speed along the waveguide, that can be
caused by internal waves, mesoscale eddies, ocean fronts or something else,
is superimposed.

In the first papers on this topic \cite{Kras,Pal,Ufn}, extremal sensitivity
of ray trajectories to the initial conditions --- ray dynamical chaos
--- has been found in simplified models of the waveguide.
In a number of recent publications \cite{Colosi,Sim,AET,Review,Wolfs} it has been 
realized that ray
chaos should play an important role in interpretating measurements
made in the long-range field experiments \cite{Duda, Worc3250} which have been
designed as a basis for ocean-acoustic tomography \cite{MunkW79, Akul}
--- determining spatio-temporal variations of the hydrological
characteristics on the real time scale from acoustical data.
The sensitivity of chaotic rays to initial conditions and small variations
of the environmental parameters causes a smearing of some timefront segments
(representing time arrivals in the time-depth plane) that has been
really observed in the field experiments \cite{Duda, Worc3250}.
Ray chaos seems to pose restrictions on the ray perturbation theory based
applications to the tomography. On the other hand, numerical experiments
\cite{Abd,Tap,Smir,Viro-web,Smir-ch} show that even at long ranges there exist 
some stable
characteristics of the sound signal which result in remarkably stable
segments of the timefront --- the main measurable characteristic in field
experiments used to reconstruct variations in the ocean environment.
In the recent paper \cite{Iomin}, maxima of the distribution function 
of the ray travel time, which lead to clustering of rays, have been analytically 
found with a simplified speed profile corresponding to a quartic oscillator.   
It has been shown in \cite{PJTF} that stable fragments of the 
timefront may correspond to regions of stability in the phase space.
Ray stability and instability are strongly influenced by the form
of the background sound speed profile.

The ray chaos studies have been especially encouraged by the field experiments
\cite{Worc3250} where acoustical signals with 75~Hz center frequency
and 37.5~Hz bandwidth, transmitted near the sound channel axis in the
eastern North Pacific Ocean, have been recorded with a vertical receiving
array between depths of 900~m and 1600~m at a range of 3250~km.
The measurements have shown a clear contrast between well-resolved
earlier portions of the received wavefronts, corresponding to steep rays
with large values of the action variable, and smearing rear segments
of the wavefronts corresponding to near-axial rays with small actions.

In this paper we study propagation of sound rays in a deep-ocean waveguide 
with  typical sound-speed profiles under internal-wave induced single-mode, 
random-like, and mixed perturbations with the aim to explain and describe 
peculiarities 
of ray chaos and ray clustering that have been found in natural and 
numerical experiments.  
The paper is organized as follows. In Sec.~\ref{secEqs} we give a brief description
of the Hamiltonian formalism in terms of the depth-momentum and
action-angle canonical variables. In Sec.~\ref{secSol} we design analytically
a background sound-speed profile, modelling typical natural deep-ocean
profiles. We integrate in quadratures the ray equations
of motion in the range-independent environment and derive exact
expressions for the angle and action variables in terms of the
depth-momentum variables. Based on the designed profile,
we consider two models of sound propagation. In Model~1 (Sec.~\ref{ssecSolM1})
we exclude from consideration the rays interacting with the ocean
surface which cannot propagate over large distance because
the profile parameters are chosen in such a way that practically
all of them interact with lossy ocean bottom as well. Shifting
the Profile 1 upwards, we obtain Model~2 with rays that may interact
with the ocean surface without interacting with the bottom.
In Sec.~\ref{secTimes} we derive analytical   expressions for ray arrival 
times and
timefronts at a given range with our model range-independent waveguides
which should be compared with those in a range-dependent waveguide.

Section~\ref{secSingle} contains results of numerical simulation with Models 1 and 2
in the presence of a single-mode perturbation induced by an internal wave.
We construct Poincar\'e maps in the polar action-angle variables
which show chains of regular islands (corresponding to different
ray-medium nonlinear resonances) surrounded by a chaotic sea.
A new insight into the phase-space structure is provided by 
plots which show by color modulation, respectively, values of variations
of the action per ray cycle length and values of the range where rays
interact with the bottom in terms of initial values of the action
and angle variables. In the end of this section we demonstrate
the possibility of determining the wavelength of the perturbation
from arrival time distribution under conditions of ray chaos
with our model profiles and the Munk canonical one.

In Sec.~\ref{secNoise} we study ray motion under a multiplicative 
noisy-like perturbation
modelling sound-speed fluctuations caused by a spectrum of internal
waves with flat and decreasing (with the wave number as $k^{-2}$)
spectral densities. We show that some rays may form {\it coherent clusters} 
consisting
of fans of rays propagating over long distances with close dynamic
characteristics. The respective plots of variations of the action
are used to clarify a mechanism of appearing coherent clusters
in local zones of stability
in the system's phase space that can survive even under a noisy-like
perturbation.
The clusterization results in appearing prominent peaks in 
arrival-time distribution functions and manifests itself in
timefronts of arriving signals as sharp strips on a smearing
background and in plots presenting ray travel time versus starting
momentum as ``shelf''-like segments.

\section{\label{secEqs}Hamiltonian equations of ray motion in an underwater 
acoustic waveguide}

Consider a two-dimensional underwater acoustic waveguide in the deep
ocean with the sound speed $c$ being smooth function of depth $z$ and
range $r$. In the geometrical-optics limit, one-way sound ray
trajectories satisfy the canonical Hamilton equations \cite{LL}
\begin{equation}
\dfrac{dz}{dr}=\dfrac{\partial H}{\partial p},\qquad
\dfrac{dp}{dr}=-\frac{\partial H}{\partial z},
\label{sys}
\end{equation}
with the Hamiltonian
\begin{equation}
H=-\sqrt{n^2(z,\,r)-p^2},
\label{1ham}
\end{equation}
where $n(z,\,r)=c_0/c(z,\,r)$ is the refractive index,
$c_0$ is a reference sound speed, $p=n\sin{\phi}$ is the analog to mechanical 
momentum, and $\phi$ is a ray grazing angle.
Only those rays that propagate at comparatively small grazing angles
can survive in the ocean at long distances, the other ones attenuate
rapidly interacting with the lossy ocean bottom.
In the paraxial approximation, the Hamiltonian can be written
in a simple form as a sum
of the range-independent and range-dependent parts \cite{Ufn}
\begin{equation}
H=H_0+H_1(r)
\label{hamsum}
\end{equation}
with the terms
\begin{equation}
H_0=-1+\dfrac{p^2}{2}+\dfrac{\Delta c(z)}{c_0},\quad
H_1=\dfrac{\delta c(z,\,r)}{c_0},
\label{ham01}
\end{equation}
where $\Delta c(z)=c(z)-c_0$, $\delta c(z,\,r)$ describes variations
of the sound speed along the waveguide. In deriving Eqs.~(\ref{ham01}),
we used the condition $|n^2(z,\,r)-1|\ll 1$, that is valid with
natural underwater sound channels, and the approximation
$n^2(z,\,r)-1\simeq-2\,\Delta c(z)/c_0$.
Moreover, in the paraxial approximation the expression
$p\simeq\tan{\phi}$ is valid.
After making the canonical transformation from the variables $(p,\,z)$
to the action--angle variables $(I,\,\vartheta)$, the Hamiltonian
may be written in the convenient form
\begin{equation}
H=H_0(I)+H_1(I,\,\vartheta,\,r).
\label{ham}
\end{equation}
The action variable is defined as the integral \cite{LL}
\begin{equation}
I=\frac{1}{2\pi}\oint p\,dz=\frac{1}{\pi}
\int\limits_{z_\text{min}}^{z_\text{max}}\sqrt{2\left(1+H_0-\frac{\Delta
c(z)}{c_0}\right)}\,dz,
\label{action}
\end{equation}
with $z_\text{min}$ and $z_\text{max}$ being the depths of the upper and
lower ray turning points, respectively. The angle variable is defined as
follows:
\begin{equation}
\vartheta=\dfrac{\partial G}{\partial I}=\left\{
\begin{aligned}
&\omega\int\limits_{z_\text{min}}^z \dfrac{dz}{p}, &p>0,\\
-&\omega\int\limits_{z_\text{min}}^z \dfrac{dz}{p}, &p\le 0,
\end{aligned}\right.
\label{angl}
\end{equation}
where $\omega$ is the angular frequency of spatial path oscillations, and
\begin{equation}
G=\int\limits_{z_\text{min}}^z p\,dz
\label{shac}
\end{equation}
is the generating function.

In a range-independent waveguide the sound-speed profile does not depend on
the range $r$. In such a waveguide the Hamiltonian $H_0$ remains constant 
along the ray trajectory, and the ray equations in the action-angle variables 
are trivial 
\begin{equation} 
\dfrac{dI}{dr}=-\frac{\partial H_0}{\partial \vartheta}=0,\qquad
\dfrac{d\vartheta}{dr}=\frac{\partial H_0}{\partial I}=\omega(I), 
\label{1} 
\end{equation} 
with the solution 
\begin{equation} 
I=I_0,\quad
\vartheta=\vartheta_0 +\omega(I_0)\,r,
\label{2} 
\end{equation} 
where $I_0=I(r=0)$ and $\vartheta_0=\vartheta(r=0)$ are initial values of 
the action and angle, respectively. In a range-independent waveguide ray 
trajectories are periodic curves. In a range-dependent waveguide 
the Hamiltonian equations in terms of 
the action and angle variables take the form \cite{Ufn}
\begin{equation}
\dfrac{dI}{dr}=-\frac{\partial H_1}{\partial \vartheta},\qquad
\dfrac{d\vartheta}{dr}=\omega+\frac{\partial H_1}{\partial I}.
\label{dot_can}
\end{equation}
The action now does not conserve along the ray path. The equations 
(\ref{dot_can}) are, in general, nonintegrable and are known to have 
chaotic solutions even under a periodic perturbation $H_1$ 
\cite{Kras,Pal,Ufn,Tap,Dan}. 

\section{\label{secSol}Exact solutions with model range-independent waveguides}

\begin{figure}[!htb]
\centerline{\includegraphics[width=\pictwidth,clip]{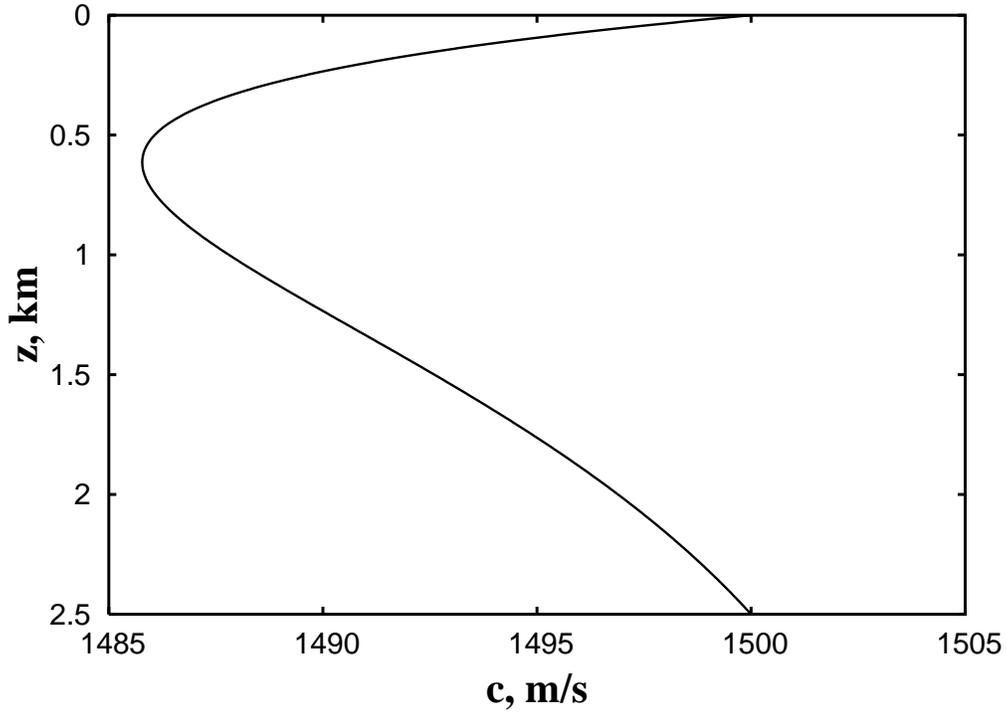}}
\caption{Analytic unperturbed sound-speed
profile referred as Profile~1 (or Model~1).}
\label{fig1}
\end{figure}

In this section we study ray nonlinear dynamics in range-independent 
waveguides with model sound-speed profiles we have designed analytically. 
Our model profiles seem to be attractive by two reasons: they are typical 
in shape for natural 
deep-ocean background sound-speed profiles and provide analytical solutions 
to the ray equations including exact expressions for the action-angle variables 
in terms of the depth-momentum variables and analytical ones for 
timefronts and ray travel times.
The model profile, hereafter referred as Profile~1 (or Model~1), is depicted 
in Fig.~\ref{fig1}.
Practically, all the rays, propagating in the corresponding waveguide, that
interact with the ocean surface interact with the ocean bottom as well.
Because of the strong attenuation of sound in the bottom, we will exclude such
exceptional rays from consideration in numerical simulation. Shifting 
Profile~1 upwards to some distance, as it is shown in Fig.~\ref{fig2}, we obtain
Profile~2 (or Model~2) with rays that may interact with the ocean surface 
without interacting with the ocean 
bottom. Both the models will be considered because some characteristics of
rays, propagating in the respective waveguides, may differ. 
The Munk canonical profile, widely used in underwater acoustics, is 
shown in Fig.~\ref{fig2}b for comparison. 

\subsection{\label{ssecSolM1}Model~1 without reflections of rays from the ocean surface}

In Ref.~\cite{Dan} we have introduced a background sound-speed profile,
shown in Fig.~\ref{fig1}, that models sound propagation through the deep
ocean

\begin{equation}
c(z)=c_0\left[1-\dfrac{b^2}{2}\,(1-e^{-az})(e^{-az}-\gamma )\right],
\qquad 0\le z\le h,
\label{prof}
\end{equation}
where $\gamma =\exp(-ah)$, $h$ is the lower border of the underwater sound channel
that is the ocean bottom,
$a$ and $b$ are adjusting parameters.
In simulation with Model~1, we used the following values of the parameters:
$a=1.0$~km$^{-1}$, $b=0.3$, $h=2.5$~km, and $c_0=1500$~m/s.
The depth of the channel axis, where the speed of sound $c(z_a)$ is minimal,
is given by
\begin{equation}
z_a=\frac{1}{a}\,\ln
{\frac{2}{1+\gamma}},
\label{ax}
\end{equation}
and the parameter $b$ is connected with $c(z_a)$ as follows: 
\begin{equation}
b=\dfrac{8}{1-\gamma}\,\sqrt{1-\frac{c(z_a)}{c_0}}.
\label{beta}
\end{equation}
The cycle length of the ray path in the channel is given by
\begin{equation}
D=2\int\limits_{z_\text{max}}^{z_\text{min}}\frac{dz}{p}
=\frac{2\pi}{a\sqrt{\gamma b^2-2E}},
\label{D}
\end{equation}
where $E=H_0+1$.
The Hamilton equations with the range-independent channel (\ref{prof})
\begin{equation}
\begin{aligned}
\dfrac{dz}{dr}&=p,\\
\dfrac{dp}{dr}&=-\frac{1}{2}ab^2e^{-az}(1+\gamma-2e^{-az})
\label{sys0}
\end{aligned}
\end{equation}
can be solved exactly
\begin{equation}
z(r)=\dfrac{1}{a}\ln
{\dfrac{a^2b^2\Bigl[1+\gamma-Q\cos{(\omega r+\vartheta_0)}\Bigr]}
{2\omega^2}},
\label{zr}
\end{equation}
\begin{equation}
p(r)=\dfrac{\omega Q\sin{(\omega r+\vartheta_0)}}{a\Bigl[1+
\gamma-Q\cos{(\omega r+\vartheta_0)}\Bigr]},
\label{pr}
\end{equation}
where $\omega=2\pi/D$ and $\vartheta_0$ are the frequency and the initial
phase of spatial oscillations of ray path in the channel, respectively.
We used the short notation in the solutions (\ref{zr}) and (\ref{pr})
\begin{equation}
Q(E)=\sqrt{(1-\gamma)^2+\frac{8E}{b^2}}.
\label{Q}
\end{equation}
The initial phase with a point source, placed at the channel axis, is
\begin{equation}
\vartheta_0=\pm\frac{\pi}{2}\mp\arcsin{\frac{Q}{1+\gamma}}.
\label{theta0}
\end{equation}
The unperturbed separatrix is defined by the value $E=0$. It is a trajectory 
that separates propagating rays touching and not touching the bottom.
Calculating the canonical variables (\ref{action}) and (\ref{angl}) with 
Model~1, we find the action
\begin{equation}
I=\frac{b}{a}\left(\frac{1+\gamma}{2}-\sqrt{\gamma-\frac{2E}{b^2}}
\right)
\label{action1}
\end{equation}
and the angle
\begin{equation}
\begin{aligned}
\vartheta=\pm
\frac{\pi}{2}\mp\arcsin{\frac{1+\gamma-(2\gamma-4E/b^2)\,e^{az}}{Q}}.
\end{aligned}
\label{angle}
\end{equation}
The old canonical variables are the following functions of the new ones:
\begin{equation}
z(I,\,\vartheta)=\dfrac{1}{a}\ln
{\dfrac{a^2b^2\,\left(1+\gamma-Q(I)\cos{\vartheta}\right)}{2\omega^2(I)}},
\label{zI}
\end{equation}
\begin{equation}
p(I,\,\vartheta)=\dfrac{\omega(I)\,Q(I)\sin{\vartheta}}{a\left(1+\gamma-Q\cos{\vartheta}\right)},
\label{pI}
\end{equation}
where
\begin{equation}
Q(I)=2\sqrt{\frac{(1+\gamma)\,aI}{b} -\frac{a^2I^2}{b^2}}, 
\label{QI}
\end{equation}
and the frequency of spatial oscillations is given by
\begin{equation}
\omega(I)=\frac{ab\,(1+\gamma)}{2} - a^2I.
\label{wI}
\end{equation}
The maximal (at $I=0$) and minimal (at $E=0$) values of the frequency $\omega(I)$
define the minimal and maximal ray cycle lengths, respectively
\begin{equation}
D_\text{min}=\frac{4\pi}{ab\,(1+\gamma)}, \qquad
D_\text{max}=\frac{2\pi}{ab\,\sqrt{\gamma}}.
\label{Dmin}
\end{equation}
The derivative $d\omega/dI$ is known as a parameter characterizing
some nonlinear properties of a sound speed profile
\begin{equation}
\frac{d\omega}{dI}=-a^2.
\label{dwi}
\end{equation}
The Hamiltonian can now be written as a function of the action variable only
\begin{equation}
H_0(I)=\frac{(1+\gamma)\,ab}{2}\,I+1-
\frac{b^2(1-\gamma)^2}{8} -\frac{a^2}{2}I^2.
\label{H0_I}
\end{equation}

\subsection{Normal mode amplitudes of the acoustic field in terms of 
ray quantities}

In this section we derive an exact analytical expression for normal mode 
amplitudes of the acoustical wave field in the range-independent waveguide 
(\ref{prof}) in terms of ray variables whose exact solutions we have found 
above. The connection between ray and modal expansions of wave fields in 
the range-independent environment is well known \cite{Brekh}. 
The normal modes of the unperturbed problem satisfy the wave equation
\begin{equation}
\frac{1}{2}\dfrac{d^2\psi_m}{dz^2}+k^2\left[E_m-
\frac{\Delta c(z)}{c_0}\right]\psi_m=0,
\label{Sturm-L}
\end{equation}
where $k=2\pi\Omega/c_0$ is the wave number in the reference medium
with the sound speed $c_0$, $\Omega$ is a carrier frequency,
and $E_m=1+H_0(I_m)$. The eigenfunctions $\psi_m(z)$, which represent normal 
modes in a range-independent waveguide, are supposed to be orthogonal and 
normalized. They constitute a complete set of basic functions in expanding 
an arbitrary wave field. 

In the Wentzel--Kramers--Brillouin approximation, the eigenvalues of the 
action variable $I_m$, corresponding to the $m$-th mode, are determined 
by the Bohr--Sommerfeld quantization rule
\begin{equation}
kI_m=m+\frac{1}{2}.
\label{I_m}
\end{equation}
The $m$-th eigenfunction $\psi_m(z)$ between its turning points 
can be represented as follows:
\begin{equation}
\psi_m(z)=\psi^{+}_m(z)+\psi^{-}_m(z),
\label{phi0}
\end{equation}
where
\begin{equation}
\psi^{\pm}_m(z)=A_m\exp{\bigl[\pm i(kG_m(z)-\pi/4)\bigr]}.
\label{phi}
\end{equation}
The phase factor is given by 
\begin{equation}
G_m(z)=G(z,\,I_m)=\int\limits_{z_\text{min}}^zp_m(z)\,dz.
\label{Sdef}
\end{equation}
The $m$-th eigenvalue $p_m$ with the model profile (\ref{prof}) can be 
easily found from Eq.~(\ref{ham01}) to be 
\begin{equation}
p_m(z)=\sqrt{2E_m+b^2(1-e^{-az})(e^{-az}-\gamma)}.
\label{p_m}
\end{equation}
The integral (\ref{Sdef}) can be calculated exactly
\begin{multline}
G_m(z)=\frac{\pi I_m}{2}+\frac{b\,(1+\gamma)}{2a}
\arcsin{\dfrac{1+\gamma -2\,e^{-az}}{Q_m}}+
\\+
\sqrt{\gamma-\frac{2E_m}{b^2}}
\arcsin{\dfrac{(1+\gamma)\,b^2-(2\gamma b^2-4E_m)
\,e^{-az}}{b^2Q_m}}-
\frac{p_m(z)}{a},
\label{Sm}
\end{multline}
where $Q_m=Q(I_m)$. The amplitude of the $m$-th mode function is given by
the following exact expression: 
\begin{equation}
A_m(z)=\sqrt{\frac{ab\,(1+\gamma)-2a^2I_m}{4\pi p_m(z)}}.
\label{Am}
\end{equation}

\subsection{\label{ssecSolM2r}Model~2 with rays reflecting from the ocean surface}

\begin{figure}[!htb]
\centerline{\includegraphics[width=\pictwidth,clip]{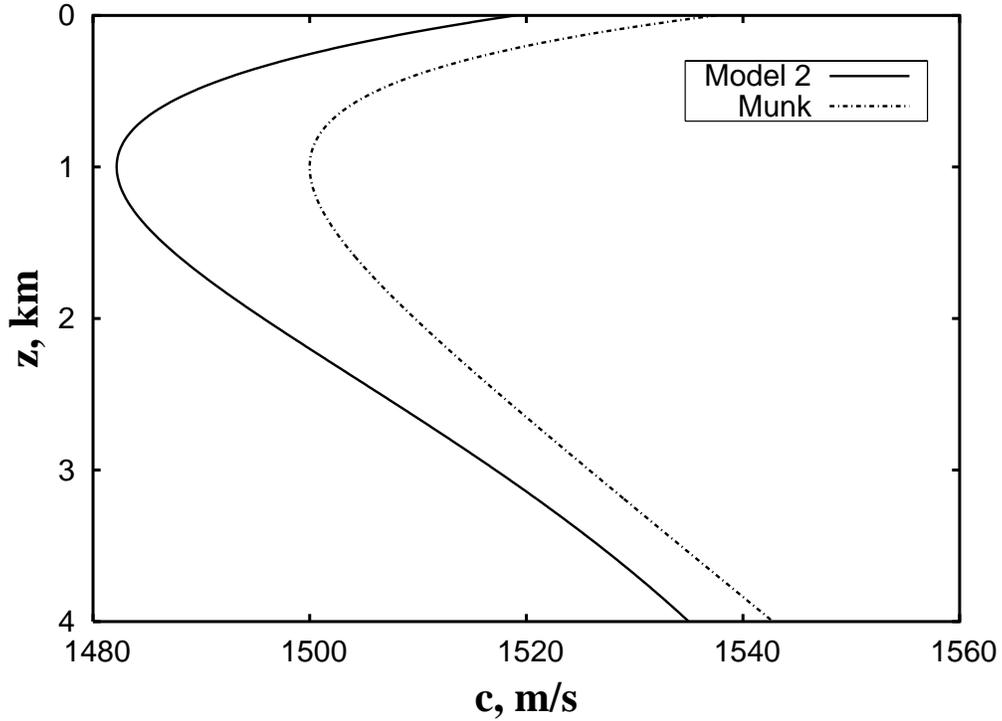}}
\caption{Analytic unperturbed sound-speed profile referred as Profile 2
(or Model~2) 
and the Munk canonical profile.}
\label{fig2}
\end{figure}

By shifting Profile~1 (see Eq.~(\ref{prof})) upward to a distance $d$, we get Profile 2 
(or Model~2) depicted in Fig.~\ref{fig2} as a solid curve 
\begin{equation}
c(z)=c_0\left[1-\dfrac{b^2}{2}(1-e^{-a(z+d)})(e^{-a(z+d)}-\gamma )\right],
\quad 
0\le z\le h,
\label{prof2}
\end{equation}
where $\gamma =\exp\bigl[-a(h+d)\bigr]$, $h$ is the maximal depth of the ocean,
$c_0=c(h)$, $a$ and $b$ are adjusting parameters. In simulation with Model~2
we have used the following values of the parameters: $a=0.5$~km$^{-1}$, $b=0.6$,
$h=4.0$~km, $d=0.15$~km, and $c_0=1535$~m/c. In contrary to Model~1, there exist in 
Model~2 rays which may interact with the ocean surface without interacting 
with the ocean bottom. We will take such rays into consideration because they 
can propagate to long distances in the ocean. For the surface-bounce 
rays $H>H_r$, where $H_r$ is given
by
\begin{equation}
H_{r}=-1-\frac{b^2}{2}(1-e^{-ad})(e^{-ad}-\gamma).
\label{Hcr}
\end{equation}
The cycle length of a ray, reflecting from the ocean surface,
is the following:
\begin{equation}
D_r=\frac{2}{a}\,\frac{\pi-\vartheta_r}{\sqrt{\gamma b^2-2E}},
\label{Dref}
\end{equation}
where we used the notation
\begin{equation}
\vartheta_r=\frac{\pi}{2}-\arcsin\left[\frac{1+\gamma-(2\gamma-4E/b^2)\,e^{ad}}{Q}\right]
\label{theta_r}.
\end{equation}
\begin{figure}[!htb]
\centerline{\includegraphics[width=\pictwidth,clip]{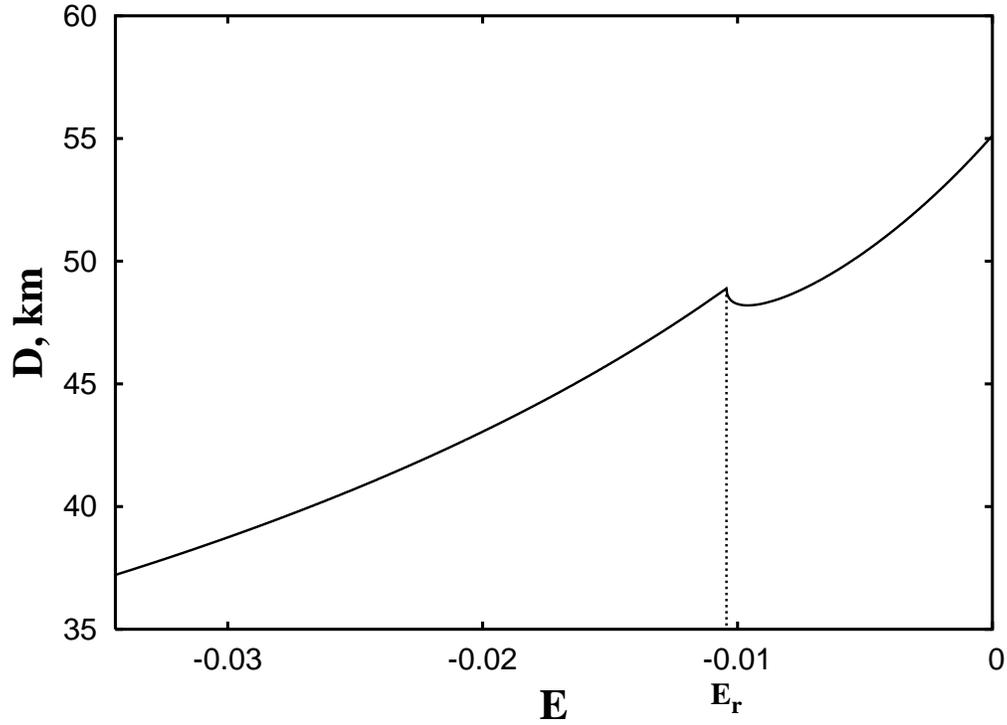}}
\caption{The cycle length of the ray path $D$ versus the ``energy'' $E$
for the unperturbed Profile~2.}
\label{fig3}
\end{figure}
In Fig.~\ref{fig3} we show the dependence of the ray cycle length $D$ on 
the ``energy'' $E=H_0+1$. The respective derivative $dD/dE$ has a 
singularity at $E_{r}=1+H_{r}$.
As in Model~1, the value $E=0$ defines the unperturbed separatrix.
We were able to find exact expressions for the action
\begin{equation}
I=\frac{p(z=0)}{\pi a}+\frac{b}{a}
\left(
\frac{1+\gamma}{4}-\frac{1+\gamma}{2\pi}
\arcsin{\dfrac{1+\gamma-2e^{-ad}}{Q}}
-\frac{\pi-\theta_r}{\pi}
\sqrt{\gamma-\frac{2E}{b^2}}
\right)
\label{action2}
\end{equation}
and the angle
\begin{equation}
\vartheta=
\left\{\begin{aligned}&
\frac{\pi}{\pi-\vartheta_r}
\left[
\frac{\pi}{2}-\vartheta_r-
\arcsin{\dfrac{1+\gamma-(2\gamma-4E/b^2)\,e^{a(z+d)}}{Q}}
\right],&
p\ge 0,
\\
&
\frac{\pi}{\pi-\vartheta_r}
\left[
\frac{\pi}{2}+
\arcsin{\dfrac{1+\gamma-(2\gamma-4E/b^2)\,e^{a(z+d)}}{Q}}
\right],&
p<0,
\end{aligned}\right.
\label{angl_ref}
\end{equation}
for the surface-bounce rays. Under reflections, the ray momentum 
is given by
\begin{equation}
p(z=0)=\sqrt{2E+b^2(1-e^{-ad})(e^{-ad}-\gamma)}.
\label{p0}
\end{equation}
The depth-momentum canonical variables in the range $H>H_r$
are the following functions of the action-angle variables:
\begin{equation}
z(I,\,\vartheta)=
\left\{\begin{aligned}&
\dfrac{1}{a}\ln
\dfrac{a^2b^2\left[1+\gamma
+Q\cos{\left(\frac{\pi}{\pi-\vartheta_r}(\vartheta+\pi)\right)}\right]}
{2\omega^2},
&
-\pi\le\vartheta\le 0, 
\\
&
\dfrac{1}{a}\ln
\dfrac{a^2b^2\left[1+\gamma
-Q\cos{\left(\frac{\pi}{\pi-\vartheta_r}\vartheta+\vartheta_r\right)}\right]}
{2\omega^2},
&
0\le\vartheta\le\pi.
\end{aligned}\right.
\label{zI2}
\end{equation}
\begin{equation}
p(I,\,\vartheta)=
\left\{\begin{aligned}&
\dfrac{\omega Q\sin{\left(\frac{\pi}{\pi-\vartheta_r}(\vartheta+\pi)\right)}}
{a\left[1+\gamma
+Q\cos{\left(\frac{\pi}{\pi-\vartheta_r}(\vartheta+\pi)\right)}\right]},
&
-\pi\le\vartheta\le 0,
\\
&
\dfrac{\omega Q\sin{\left(\frac{\pi}{\pi-\vartheta_r}\vartheta+\vartheta_r\right)}}
{a\left[1+\gamma
-Q\cos{\left(\frac{\pi}{\pi-\vartheta_r}\vartheta+\vartheta_r\right)}\right]},
&
0\le\vartheta\le\pi,
\end{aligned}\right.
\label{pI2}
\end{equation}
where $\omega$ is given by the same formula as in Model~1
\begin{equation}
\omega=a\sqrt{\gamma b^2-2E}.
\label{omega}
\end{equation}

As to normal modes of the unperturbed waveguide (\ref{prof2}), they satisfy
the respective wave equation (\ref{Sturm-L}) with the Bohr-Sommerfeld
quantization rule 
\begin{equation}
kI_m=\left\{
\begin{aligned}
&m+\frac{1}{2},\quad &H\le H_r,\\
&m-\frac{1}{4},\quad &H>H_r.
\end{aligned}
\right.
\label{I_m2}
\end{equation}
At $H\le H_r$, the phase factor $G_m$ and the amplitude $A_m(z)$ 
of the $m$-th mode function are given 
by Eqs.~(\ref{Sm}) and (\ref{Am}), respectively.
At $H > H_r$, we get
\begin{multline}
G_m=\frac{b}{a}\,\sqrt{\gamma-\frac{2E_m}{b^2}}
\left[
\arcsin{\dfrac{1+\gamma-\left(2\gamma-4E_m/b^2\right)\,e^{a(z+d)}}{Q_m}}+
\vartheta_r(E_m)-\frac{\pi}{2}
\right]+\\
+\frac{(1+\gamma)\,b}{2a}
\left[\arcsin{\dfrac{1+\gamma-2e^{-a(z+d)}}{Q_m}}-
\arcsin{\dfrac{1+\gamma-2e^{-ad}}{Q_m}}\right]+
\frac{p_m(z=0)-p_m(z)}{a}.
\label{Sm2}
\end{multline}
The amplitude of the $m$-th mode function at $H>H_r$ is
\begin{equation}
A_m(z)=\sqrt{\frac{ab(1+\gamma)-2a^2I_m}{4(\pi-\vartheta_r)\,p_m(z)}}.
\label{Am2}
\end{equation}

\section{\label{secTimes}Ray arrival times and timefronts in range-independent and 
range-dependent waveguides}

Internal waves in the ocean induce lateral variations of the sound speed.
As a result, the ray cycle length and the ray action are not invariants as 
in range-independent waveguides
but vary slowly along the ray path.
Even very small variations of the sound speed may cause under typical
conditions exponential divergence of rays with initially close grazing angles, 
the phenomenon known as ray chaos \cite{Ufn}.
The model of a ``frozen'' medium is usually adopted, 
where one may neglect temporal variations in the environment and take into 
account only its spatial variations
due to comparatively small propagation time of sound in the ocean.
Then variations of the speed of sound may be described by the expression
\begin{equation}
\delta c(z,\,r)=\delta c_\text{rms}(z)\,\xi(z,\,r),
\label{dc0}
\end{equation}
where $\delta c_\text{rms}$ is the root-mean-square value of sound-speed
fluctuations. Following to Refs. \cite{Smith1,Smir}, we shall describe the
fluctuations by the simple formula
\begin{equation}
\delta c_\text{rms}(z)=\varepsilon c_0\frac{z}{B}\,e^{-2z/B},
\label{dc}
\end{equation}
where $\varepsilon$ is a measure of the strength of the range-dependent perturbation and $B$, the termocline
depth scale, is chosen to be 1 km. Throughout the paper, we use the 
perturbation models with only
longitudinal modes of internal waves, i.~e., $\xi(z,\,r)=\xi(r)$. Then the
perturbed Hamiltonian may be written as follows:
\begin{equation}
H=H_0(I)+\varepsilon V(I,\,\vartheta)\,\xi(r).
\label{ham_xi}
\end{equation}
Let us represent the perturbation in the form of the Fourier series over the 
cyclic variable $\vartheta$
\begin{equation}
V(I,\,\vartheta)=\frac{1}{2}\sum_{m=1}^\infty
V_{m}(I)\,e^{im\vartheta}+\text{c.c.}.
\label{v_ryad}
\end{equation}
The equations of motion are
\begin{equation}
\dfrac{dI}{dr}=-\frac{i}{2}
\sum_{m=1}^\infty mV_{m}(I)\,e^{im\vartheta}\xi(r)+\text{c.c.},
\label{di}
\end{equation}
\begin{equation}
\dfrac{d\vartheta}{dr}=\frac{2\pi}{D}+\varepsilon\frac{\partial V}{\partial I}\,\xi(r).
\label{dot_ang}
\end{equation}
The function $V(I,\,\vartheta)$ is an analytical one with the Fourier amplitudes
exponentially decreasing with increasing the number $m$. With Model~1 and
perturbation (\ref{dc}), it has the form
\begin{equation}
V(I,\,\vartheta)=a^{-1-4/a}\left(\frac{4\omega^4(I)}
{b^4\bigl(1+\gamma-Q(I)\cos{\vartheta}\bigr)^2}\right)^{1/a}
\ln{\dfrac{a^2b^2\left[1+\gamma-Q(I)\cos{\vartheta}\right]}{2\omega^2(I)}}.
\label{v_1}
\end{equation}

\begin{figure}[!htb]
\centerline{\includegraphics[width=\pictwidth,clip]{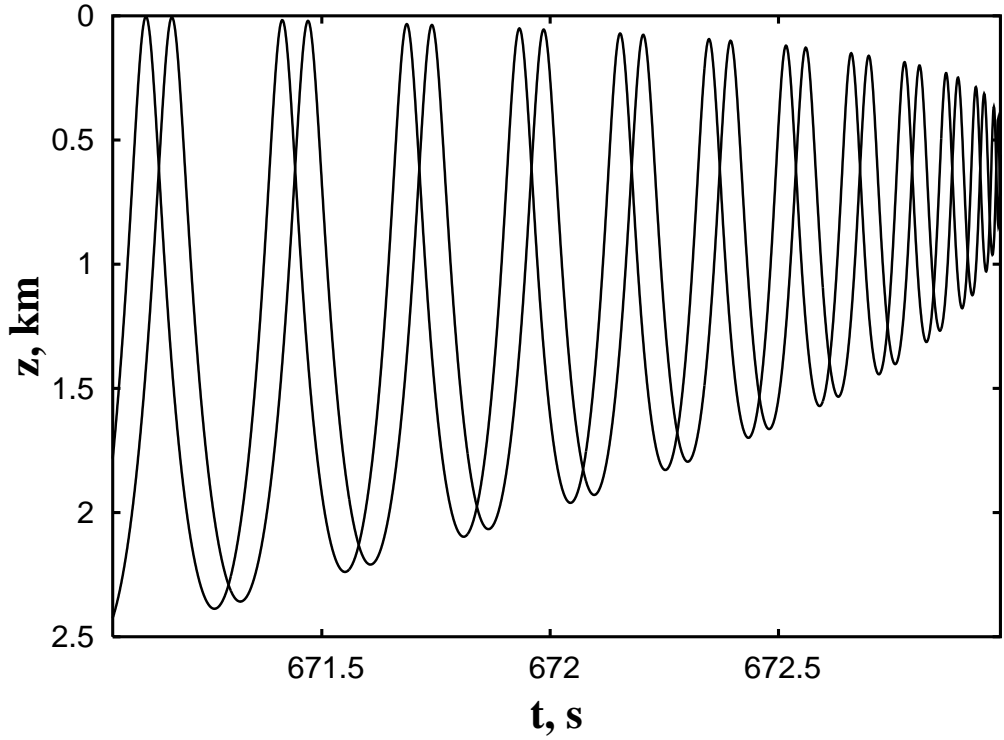}}
\caption{Timefront for the unperturbed Profile 1:
ray depth $z$ versus ray travel
time $t$ at the range 1000 km.}
\label{fig4}
\end{figure}

Methods of the acoustic tomography are actively used for studying 
spatio-temporal variations in the
ocean on the real time scale \cite{MunkW79, Akul}. When the sound waves 
propagate over long distances,
an effective means for monitoring the medium is based on the effect of 
spatial variations of the sound speed
on the signal arrival times, one of the main measurable characteristic 
in long-base acoustical experiments.
Extensive field measurements, that have been carried out in recent years 
\cite{Duda, Worc3250}, showed smearing
of timefront segments in the rear of the sound pulse. Hardly resolvable 
microfolds in the late-arriving portions of the timefront, to be observable 
in field experiments, can be reasonably explained by
the ray's sensitivity to initial conditions. Without internal waves the timefront 
has a smooth folded accordion shape due to refraction as in Fig.~\ref{fig4}. 
In the presence of internal waves, a nonuniformity of ray arrivals along the 
folded fronts appears (see Fig.~\ref{fig16}). Zooming would reveal the 
presence of microfolds along the macroscopic segments of the timefront under 
consideration. 

In accordance with the Fermat's principle, ray arrival time to 
a point $R$ along a waveguide is calculated with the help of 
the Lagrangian $L$ 
\begin{equation} 
t=\dfrac{1}{c_0}\int\limits_0^RL\, dr 
=\dfrac{1}{c_0}\int\limits_0^R (p^2-H)\,dr. 
\label{time} 
\end{equation} 
At sufficiently long ranges, $R/D\gg 1$, the Lagrangian $L$ may be 
considered as a function of the action
\begin{equation}
L(I)=2\pi \dfrac{I}{D(I)}-H_0(I).
\label{lag_i}
\end{equation}
Following to Eq.~(\ref{time}), ray arrival time to the point $R$
along a range-dependent waveguide is given by
\begin{equation}
t=\dfrac{R}{c_0}\aver{L(I)},
\label{tcom}
\end{equation}
where $\aver{\dots}$ means an averaging over $r$.
In a range-independent waveguide, arrival times for
long-range paths can be simply calculated to be
\begin{equation}
t=\dfrac{RL}{c_0} \simeq \dfrac{R}{c_0}\left(2\pi\frac{I}{D}-H_0\right), 
\label{t_un}
\end{equation}
with the Lagrangian $L$ being an invariant. With a point sound source, all the 
invariants are
functions of the initial value of the momentum $p(r=0)=p_0$, i.~e., $t=t(p_0)$.
Ray arrival time at the fixed range $R$ is maximal with axial rays (because
the sound speed is minimal at the channel axis), and decreases in average
with increasing $p_0$. If $t(p_0)$ is a monotonic function and a waveguide is
range-independent, the so-called timefront, which represent ray arrivals 
in time-depth plane, can be calculated explicitly from the equation for the
trajectory (see, for example, Eq.~(\ref{zI}) for Model~1) with the parameters
being functions of the Lagrangian $L$. 

In order to demonstrate it in Model~1
with the range-independent waveguide shown in Fig.~\ref{fig1}, we use
Eq.~(\ref{lag_i}) to find the action
\begin{equation}
I(L) = \frac{1}{a}\sqrt{2+\frac{b^2(1-\gamma)^2}{4}-2L},
\label{Y}
\end{equation}
the spatial frequency of nonlinear oscillations
\begin{equation}
\omega(L) = a\left(b\,\frac{1+\gamma}{2}-aI(L)\right),
\label{wt}
\end{equation}
and the quantity
\begin{equation}
Q(L) = 2\sqrt{\frac{(1+\gamma)\,aI(L)}{b}
-\frac{a^2I^2(L)}{b^2}},
\label{Qt}
\end{equation}
as functions of the Lagrangian $L$. After substituting 
Eqs.~(\ref{tcom}), (\ref{Y})--(\ref{Qt}) into the ray-trajectory 
equation (\ref{zI}),
we get the timefront of the sound signal in the waveguide with
the sound-speed Profile 1
\begin{equation}
z(t) \simeq \dfrac{1}{a}\ln
{\dfrac{a^2b^2\Bigl[1+\gamma-Q(t)\cos{\bigr(\omega(t)R+\vartheta_0\bigr)}\Bigr]}
{2\omega^2(t)}}.
\label{zt}
\end{equation}
The respective plot, presenting ray depths against arrival times at the
range $R=1000$~km, is shown in Fig.~\ref{fig4}.
We see a typical two-folded accordion-like structure due to refraction
with positive and negative values of grazing angle.
The ray arrivals are spread in a smooth and predictable
way with the late-arriving portion of the timefront formed by the axial rays.

In the presence of a perturbation, timefront can be computed approximately as a
sum of representing points of sound pulses
\begin{equation}
T(z,\,t) \simeq \sum_iT_i=\sum_i\delta(z-z_i)\,\delta(t-t_i),
\label{ztp}
\end{equation}
where $z_i$ and $t_i$ are the depth and arrival time of the $i$-th pulse.
With the help of Eq.~(\ref{zI}), the depth for each pulse $z$
may be written as a function of initial $I_0$, final $I_f$, and mean,
$\aver{I}$, values of the action,
which may be considered as independent variables under conditions of strong
chaos
\begin{equation}
z \simeq \dfrac{1}{a}\ln{\dfrac{a^2b^2\Bigl[1+\gamma-Q(I_f)
\cos{\bigl(\omega(\aver{I})R+\vartheta_0(I_0)\bigr)}\Bigr]}
{2\omega^2(I_f)}}.
\label{zti}
\end{equation}

Let us consider now distribution of rays over their arrival times, $f(t,\,R)$, 
at a fixed range $R$. In a range-independent waveguide, it is
determined by an initial distribution of grazing angles only.
The respective distribution function for rays,
started at small grazing angles with $\phi_0\simeq p_0$
(that may propagate over large distances), is $f_0(p_0)$. Using the condition 
of the conserved normalization 
\begin{equation}
\int f_0(t,\,R)\, dt = \int f_0(p_0)\, dp_0, 
\label{norm}
\end{equation}
we get
\begin{equation}
f_0(t,\,R)=f_0(p_0)\dfrac{dp_0}{dt}.
\label{f0}
\end{equation}
With Model~1 we get from Eqs.~(\ref{t_un}), (\ref{ham01}), (\ref{D}) and 
(\ref{action1})
\begin{equation}
f_0(t,\,R)=\frac{f_0(p_0)c_0\bigl[b\,(1+\gamma)-2U(t)\bigr]}
{RU\sqrt{b\,(1+\gamma)U(t) - U^2(t)}},
\label{f1}
\end{equation}
where the short notation is used
\begin{equation}
U(t)=\sqrt{\frac{b^2}{4}\left(1-\gamma\right)^2+2-\frac{2c_0t}{R}}.
\label{u}
\end{equation}
In a range-dependent waveguide, the distribution function of ray arrival times
is given by
\begin{equation}
f(t,\,R)=AF(t,\,R)\,f_0(p_0)\dfrac{dp_0(t)}{dt},
\label{psrs}
\end{equation}
where $A$ is a normalization constant. The function $F(t,\,R)$ describes the
effect of the range dependence of the sound speed on the distribution of ray
arrival times. It is defined, mainly, by the structure of the phase
space of the perturbed system \cite{PJTF}. We shall use the function
$F(t,\,R)$ as a convenient tool for analyzing clusterization of rays.

It should be noted that the formulas for arrival times (\ref{t_un}) and  
timefronts (\ref{zt}) and  (\ref{zti}) are approximated ones. Comparing 
Fig.~\ref{fig4}, plotted using (\ref{zt}), with numerical simulation shown in 
Fig.~\ref{fig12}a, one can see the difference between the two timefronts 
especially 
in the neighbourhoods of extrema. However, these formulas provide a correct  
general image of timefronts and give a simple analytical 
connection between measurable ray characterictics and  the phase-space ray 
variables which can be used to explain such peculiarities of timefronts 
as sharp stripes.  

\section{\label{secSingle}Ray chaos in the presence of a single-mode perturbation}
\subsection{\label{ssecSolPS}The phase-space structure}

In this section we consider a single-mode sound-speed perturbation (\ref{dc0})
with $\delta c_\text{rms}(z)$ given
by (\ref{dc}) and the horizontal dependence of the
internal-wave induced perturbation given by
\begin{equation}
\xi(r)=\cos{kr}=\cos{\frac{2\pi r}{\lambda}},
\label{perv}
\end{equation}
where $\lambda$ is the wavelength of the internal wave. The Hamilton equations
take the form
\begin{equation}
\dfrac{dI}{dr}=-\frac{i}{2}\varepsilon\sum_{l,\,m}lV_{lm}e^{im\Psi}+\text{c.c.},
\label{i}
\end{equation}
\begin{equation}
\dfrac{d\vartheta}{dr}=\omega+\frac{\varepsilon}{2}
\sum_{l,\,m}\frac{V_{lm}}{dI}\,e^{im\Psi}+\text{c.c.},
\label{th}
\end{equation}
where the new phase $\Psi=\vartheta-\dfrac{lkr}{m}+\dfrac{\phi_0}{m}$ is
introduced. Ray trajectories are captured in a ray-medium space
nonlinear resonance if the condition
\begin{equation}
m\omega(I)=lk,
\label{rescond}
\end{equation}
is satisfied with $l$ and $m$ being integers. This condition can be satisfied
at different values of the action variable
$I_\text{res}$ corresponding to resonant tori. Phase oscillations in
vicinities of the resonant tori are described by the universal Hamiltonian
of nonlinear resonance \cite{Ufn, Chir}
\begin{equation}
H_u=m\left(\frac{1}{2}\,\bigl|\omega'_I(I_\text{res})\bigr|\left(\Delta I\right)^2
+\varepsilon|V_{lm}|\cos{m\Psi}\right),
\label{universal}
\end{equation}
where $\omega'_I(I_\text{res})=d\omega(I_\text{res})/dI$. The width of the
resonance in terms of spatial frequency can be approximately estimated as
\begin{equation}
\Delta\omega=|\omega'_I|\Delta I=
2\sqrt{\varepsilon|\omega'_I|V_{lm}},
\label{shir}
\end{equation}
where $\Delta I$ is the width of the nonlinear resonance in terms of 
the action variable. In accordance with Chirikov's criterion \cite{Chir}, 
global chaos may arise if 
\begin{equation} 
\frac{\Delta\omega}{\delta\omega}\simeq 1,
\label{Chi} 
\end{equation} 
i.~e., if two nonlinear resonances, centered at $\omega$ and 
$\omega + \delta \omega$, overlap. 
Those resonances that overlap slightly
form islands in the phase space, areas of stable ray motion in a chaotic sea.
Nearby the island's borders,
one can find the so-called zones of stickiness where chaotic trajectories
may be localized for long distances $r$ \cite{Zchaos}.

\begin{figure}[!htb]
\centerline{\includegraphics[width=\pictwidth,clip]{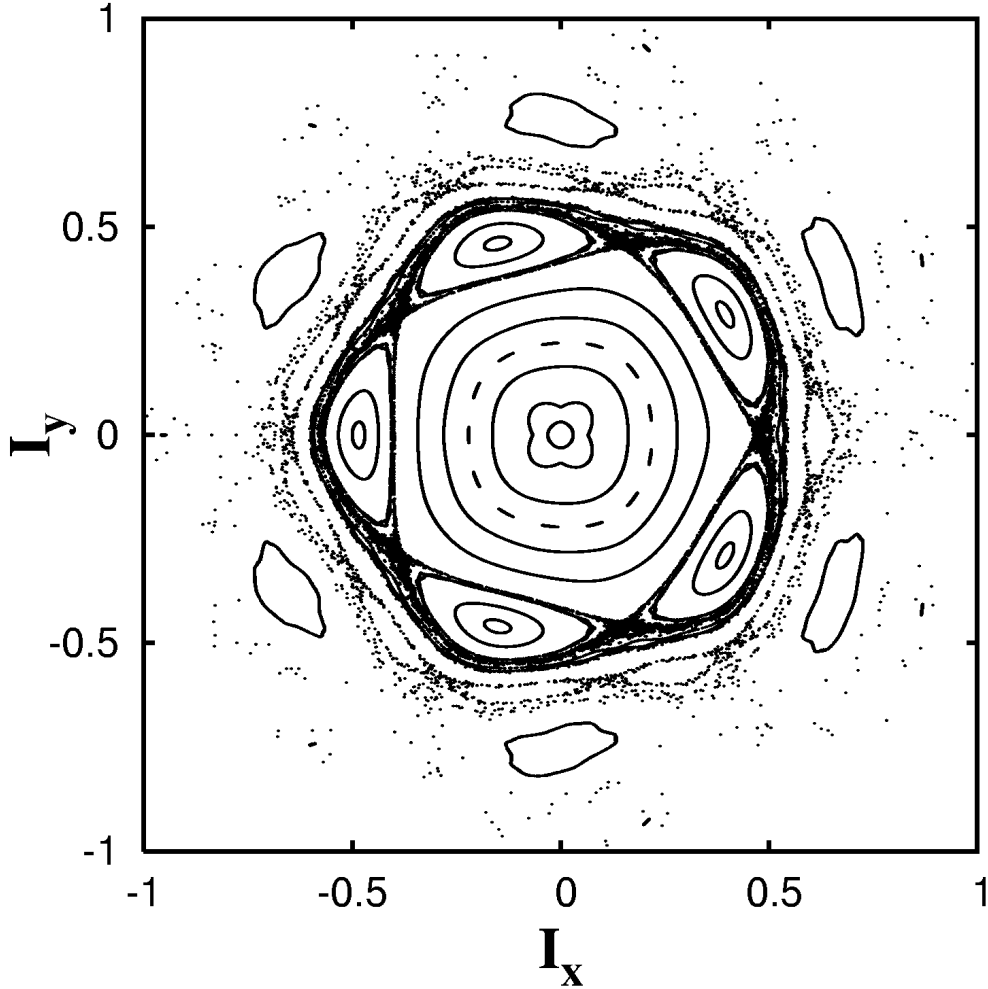}}
\caption{Poincar\'e map in the normalized polar action--angle variables for Model~1
with the parameters of the periodic perturbation, $\lambda=10$~km 
and $\epsilon=0.0025$.}
\label{fig5}
\end{figure}

To visualize the structure of the phase space, we construct a Poincar\'e 
map integrating numerically the ray equations for Model~1 with the single-mode 
perturbation (\ref{perv}). Figure~\ref{fig5} demonstrates such a map 
with the perturbation wavelength $\lambda=10$~km
and a comparatively weak perturbation strength $\varepsilon=0.0025$. 
It is a two-dimensional slice of the ray motion in a three-dimensional space 
$(I_x,\, I_y,\, r \mod \lambda)$ with $I_x=I\cos \vartheta$ and 
$I_y=I\sin \vartheta$ being the polar action-angle variables normalized 
to the separatrix value of the action $I_s$ (given by Eq.(\ref{action1}) 
at $E=0$) which is a maximal acceptable value of the action.   
 
A typical (with 
Hamiltonian systems) picture with stable islands filled with regular trajectories 
surrounded by a chaotic sea is seen in the figure. 
The chains with 5 and 6 islands correspond to the primary resonances
of the first order ($l=1$) with $m=5$ and 6, respectively. The chain
with 11 islands is located between
them and corresponds to the second-order resonance with $l=2$ and $m=11$.
Even a higher-order
resonance with 16 islands (between the $(l=1,\ m=5)$ and $(l=2,\ m=11)$ 
resonances) is seen in Fig.~\ref{fig5}. Concentration of points near
island's boundaries indicates sticky trajectories.

\begin{figure*}[!htb]
\parbox{0.49\textwidth}{{\Large\bf a)\\}
\includegraphics[width=0.49\textwidth,clip]{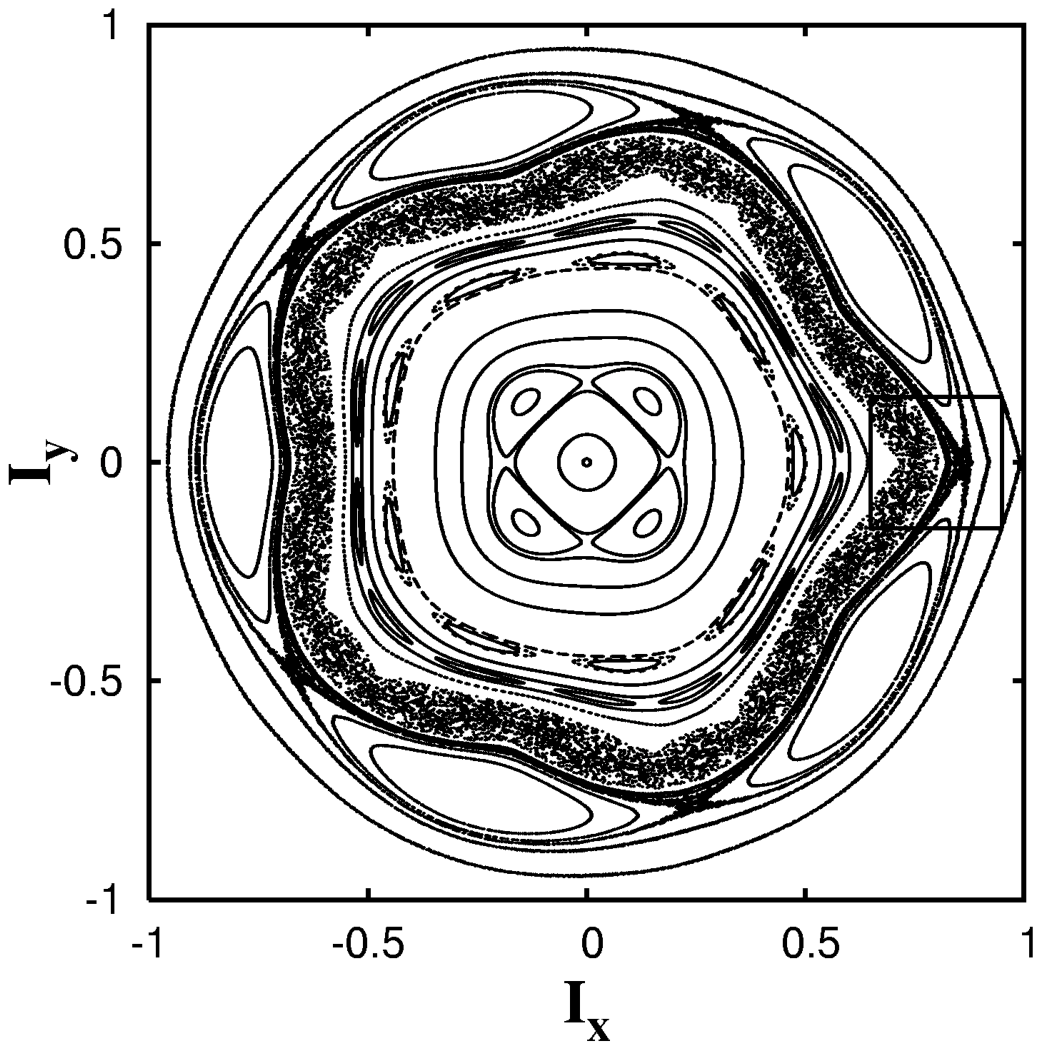}}\hfill
\parbox{0.49\textwidth}{{\Large\bf b)\\}
\includegraphics[width=0.49\textwidth,clip]{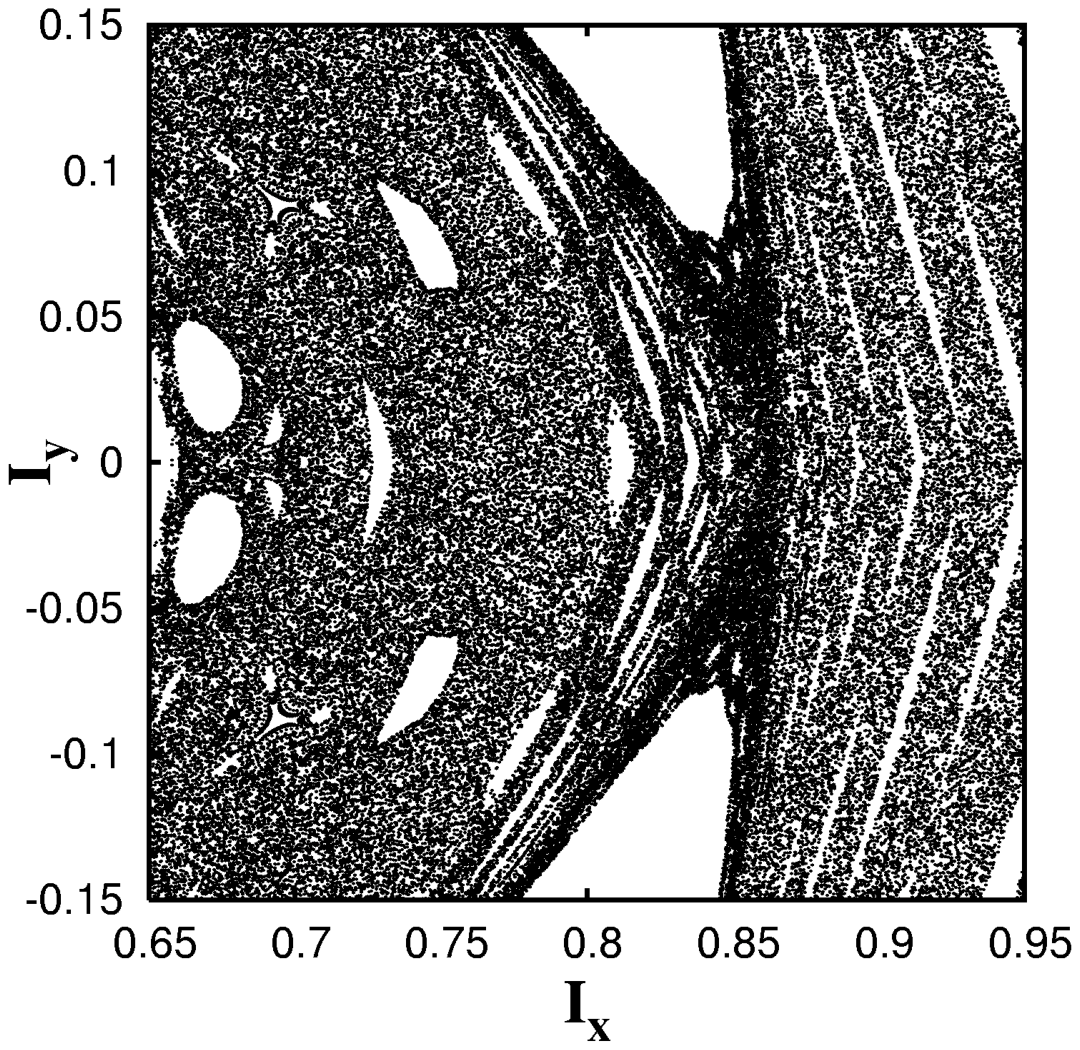}}
\caption{(a) Poincar\'e map in the polar normalized action--angle variables for Model~2
with the parameters of the periodic perturbation, $\lambda=10$~km
and $\epsilon=0.005$.
(b) Zoom of the small region of the stochastic layer indicated in (a).}
\label{fig6}
\end{figure*}

Figure~\ref{fig6} shows the Poincar\'e map with rays that may reflect 
from the ocean surface (Model~2 with $\varepsilon=0.005$ and $\lambda=10$~km). 
In difference from Model~1, a stochastic layer appears inside the separatrix 
loop in a vicinity of the critical value of the action $I_r=I(H_r)$ 
(Fig.~\ref{fig6}a). We remind that reflection of rays from the ocean surface in 
Model~2 occurs at $H > H_r$ (see Eq.~(\ref{Hcr})). The origin of the localized 
stochastic layer can be explained as follows. The distance between the 
resonances of the $m$-th and $m+1$-th orders in terms of spatial frequency is 
equal to 
\begin{equation}
\delta\omega=\frac{k}{m}-\frac{k}{m+1}\simeq\frac{\omega^2}{k}
\propto D^{-2},
\label{rasst}
\end{equation}
and decreases rapidly with increasing $D$. Since the ray cycle length $D$ 
has a local maximum at $E=E_r$ (Fig.~\ref{fig3}), the resonance overlapping 
in accordance with (\ref{Chi}) is maximal near $I(E_r)$.
This stochastic layer is isolated from the separatrix by invariant curves. 
As a result, the respective rays are trapped inside the layer forever and 
their motion is strongly influenced by the fractal microstructure of the 
stochastic layer \cite{Zas-fr, Zchaos, Zas-PhA}.
The fine structure of the phase space is demonstrated in Fig.~\ref{fig6}b 
where a zoom of the region of the stochastic layer near 
$\vartheta=0$ is shown. Chains of microislands corresponding to primary 
and secondary resonances are seen in the figure. It should be noted that an 
analogous localized stochastic layer has been found with the Munk canonical 
profile \cite{Smir, Zas-fr}.

\begin{figure}[!htb]
\centerline{\includegraphics[width=\pictwidth,clip]{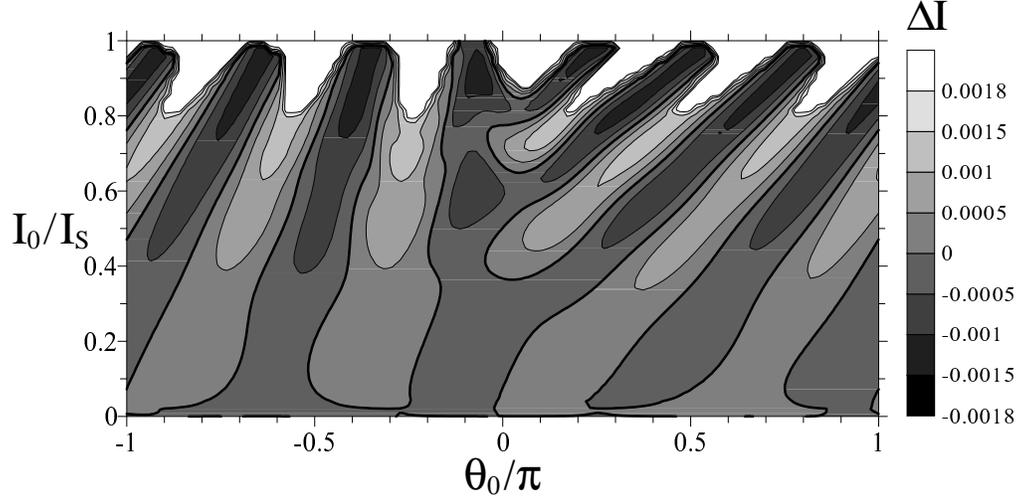}}
\caption{
Plot representing variations of the action $\Delta I$ 
per ray cycle length for Model~1 ($\lambda=10$~km and $\epsilon=0.005$) 
in the plane of the normalized initial values of the action and angle. 
Bold lines correspond to zero variations of the action per ray cycle length.}
\label{fig7}
\end{figure}

Another way to visualize the phase-space structure is provided by the plot
that shows by color modulation
values of variations of the action during the ray cycle length $\Delta I$ 
in the plane of initial values of the action and angle 
variables normalized to the separatrix value $I_s$ and $\pi$, respectively.
More exactly, $\Delta I$ is a variation of the action between to successive 
crossings of the line $\theta = {\rm const}$ by a ray. It depends on the initial 
value of the range $r_0$ and may strongly vary in the chaotic regime. 
It is a distribution of variations of the action 
over the phase space that has important physical meaning. The number of 
positive variations of the action (``hills'') and the number of its 
negative variations (``hollows'') are stable characteristics of the system 
(independent on initial conditions) describing the phase oscillations.  
The respective map for Model~1 with a
single-mode perturbation, presented in Fig.~\ref{fig7}, demonstrates an alternating
``hills'' and ``hollows'' corresponding to different values of the
phase $\Psi$. Due to the phase dependence, 
this structure periodically depends on initial values of the range 
variable $r$. The ``hills'' and
``hollows'' are separated from each other by ``zero lines'', which 
correspond to zero variations of the action per cycle length, and may be
``stable'' and ``unstable''. The ``stable'' ones transverse the elliptic points
of the potential in Eq.~(\ref{universal}),
and the ``unstable'' ones transverse the respective hyperbolic points.
It should be noted that ``hills'' and ``hollows'' may intersect
each other.

Spatial variations of the sound speed along a waveguide may cause the known
effect of ray escaping 
\cite{Kras} when some rays reach the unperturbed separatrix due to diffusion
in the action and quit the sound channel.
Those rays are supposed to quit the channel which interact with the ocean
bottom and therefore attenuate
rapidly. First of all, steep rays with comparatively small arrival times
(corresponding to higher modes of the sound field) will escape.
The escaping takes place even under an adiabatic perturbation \cite{Krav}
but it has peculiarities under the ray chaos conditions.

\begin{figure}[!htb]
\centerline{\includegraphics[width=\pictwidth,clip]{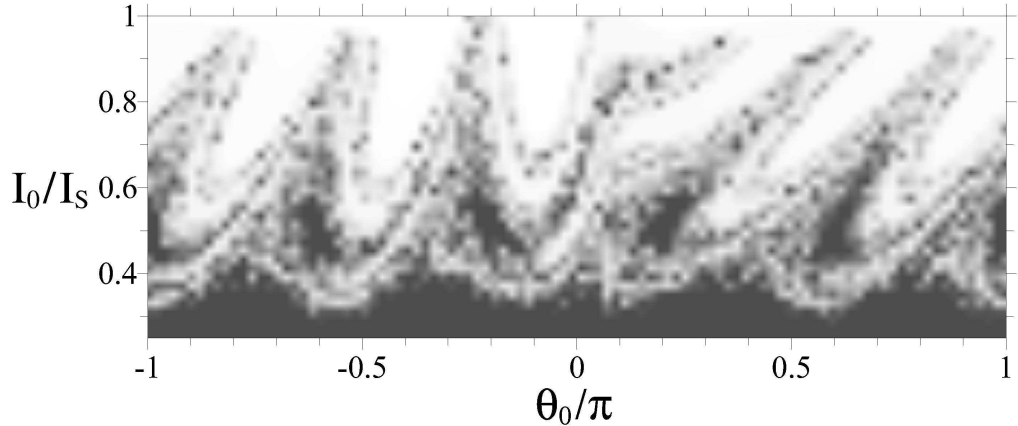}}
\caption{Plot representing the values of the range $r$, where rays interact 
with the ocean bottom, for the same model and the same parameters
as in Fig.~\ref{fig7}. White color corresponds to those rays that quit the 
channel during the first ray cycle, i.~e. at $r \leq  70$~km.}
\label{fig8}
\end{figure}
In the plot, presented for Model~1 
in Fig.~\ref{fig8}, color modulates the values
of the range $r$ where rays interact with the ocean bottom, with white color 
corresponding to rays which quit the waveguide during the first cycle, i.~e.  
at $r \leq 70$~km (which is a cycle length of a trajectory nearby 
the separatrix)  
whereas the black one corresponds to the values $r \geq 1000$~km. 
As in the case with variation of the action, distribution of the values of 
range for the rays interacting with the ocean bottom is a stable characteristic 
of the system.  
Topology
of the plot is complicated in those areas in the
$\left(I_0/I_s\right)$ --- $\left(\vartheta_0/\pi\right)$
plane which correspond to the
stochastic layer. Its patchiness reflects a complicated inhomogeneity of the
phase space. We want to stress
that channels for escaping are formed in the same places where the
``hills'' are situated in the 
plot representing variations of the action per ray cycle length 
(Fig.~\ref{fig7}).
The values of the action variable grow rapidly in such channels resulting in
increasing trajectory amplitudes. One can see in Fig.~\ref{fig8} black spots 
corresponding to those ranges of the initial conditions of ray trajectories that 
never quit the waveguide. They correspond to the respective islands 
of regular motion on the Poincar\'e map (Fig.~\ref{fig5}). The patched dark 
areas between the channels with rapidly escaping rays correspond to the initial 
conditions with long but finite lengths of escaping. 
The angular structure of the distribution of the lengths of escaping 
in the phase space is an evidence
of a non-ergodic ray diffusion. In difference from the 
plot representing variations of the action per ray cycle length, the 
plot with escaping rays has a manifested
patched structure due to fractal properties of the phase space typical for 
open chaotic Hamiltonian systems
with weak mixing (for a review see \cite{Zas-PhA}). The hierarchy of islands
and chains of islands with
sticky zones near the island's boundaries, which are repeated at all scales
of resolution, produces dynamical traps
where representing particles may be trapped for a long distance (time).
It results eventually in anomalous
diffusion and power-law distribution functions \cite{Zas-PhA}. It is
worthwhile to mention that escaping of rays is an analog of trapping (escaping) 
of chaotically advected passive
particles in open hydrodynamic flows (see, for example, \cite{Dan2}).

\subsection{\label{ssecSolD}Timefront structure under a periodic perturbation}

Let us consider now the role of the ray-medium nonlinear resonance in forming
the structure of timefronts of sound signals. Due to nonlinear resonance, 
the ray arrival
time of a sound signal along a given ray,
captured in a nonlinear resonance near the given value of the action $I_\text{res}$,
tends to its unperturbed value $t(L_\text{res})$
with increasing the distance \cite{Ufn}
\begin{equation}
t(L_\text{res})=\frac{RL(I_\text{res})}{c_0}.
\label{t_res}
\end{equation}
If the width of the resonance is sufficiently large and if there are
sufficiently many rays captured in the resonance,
the distribution function $F(t,\,R)$ (see Eq.~(\ref{psrs})) has a pronounced
peak near the value $t(L_\text{res})$. Moreover, it is possible with the help
of $F(t,\,R)$ to find the spatial period of a perturbation mode if the
arrival time is unambiguously defined by the ray cycle length \cite{PJTF}.
Inverting the resonance condition (\ref{rescond}), we get
\begin{equation}
lD_\text{res}=m\lambda.
\label{rescond1}
\end{equation}
There can be several peaks of the function $F(t,\,R)$ with large 
amplitudes corresponding to resonances with small $m$ at $l=1$. Thus, if
$F(t,\,R)$ exhibits at least two distinct peaks with comparable amplitudes,
we can determine the period of a single-mode perturbation as
\begin{equation}
\lambda=D(t_\text{res1})-D(t_\text{res2}),
\label{def_d}
\end{equation}
where $t_\text{res1}$ and $t_\text{res2}$ are arrival times corresponding
to the two peaks. In general, the values of $D(t_\text{res})$
can be calculated numerically with the help of Eq.~(\ref{t_un}).
Our model Profiles~1 and 2 admit analytical
calculation of the resonant cycle length. In Model~1, for example,
one can deduce from Eq.~(\ref{t_un}) a formula
connecting the ray cycle length $D$ with the arrival time $t$
\begin{equation}
D(t)=\frac{2\pi}{a}\left (b\,\frac{1+\gamma}{2}-\sqrt{\frac{b^2}
{4}\left(1-\gamma\right)^2+2-\frac{2c_0t}{R}}\right )^{-1}.
\label{dt}
\end{equation}

\begin{figure}[!htb]
\centerline{\includegraphics[width=\pictwidth,clip]{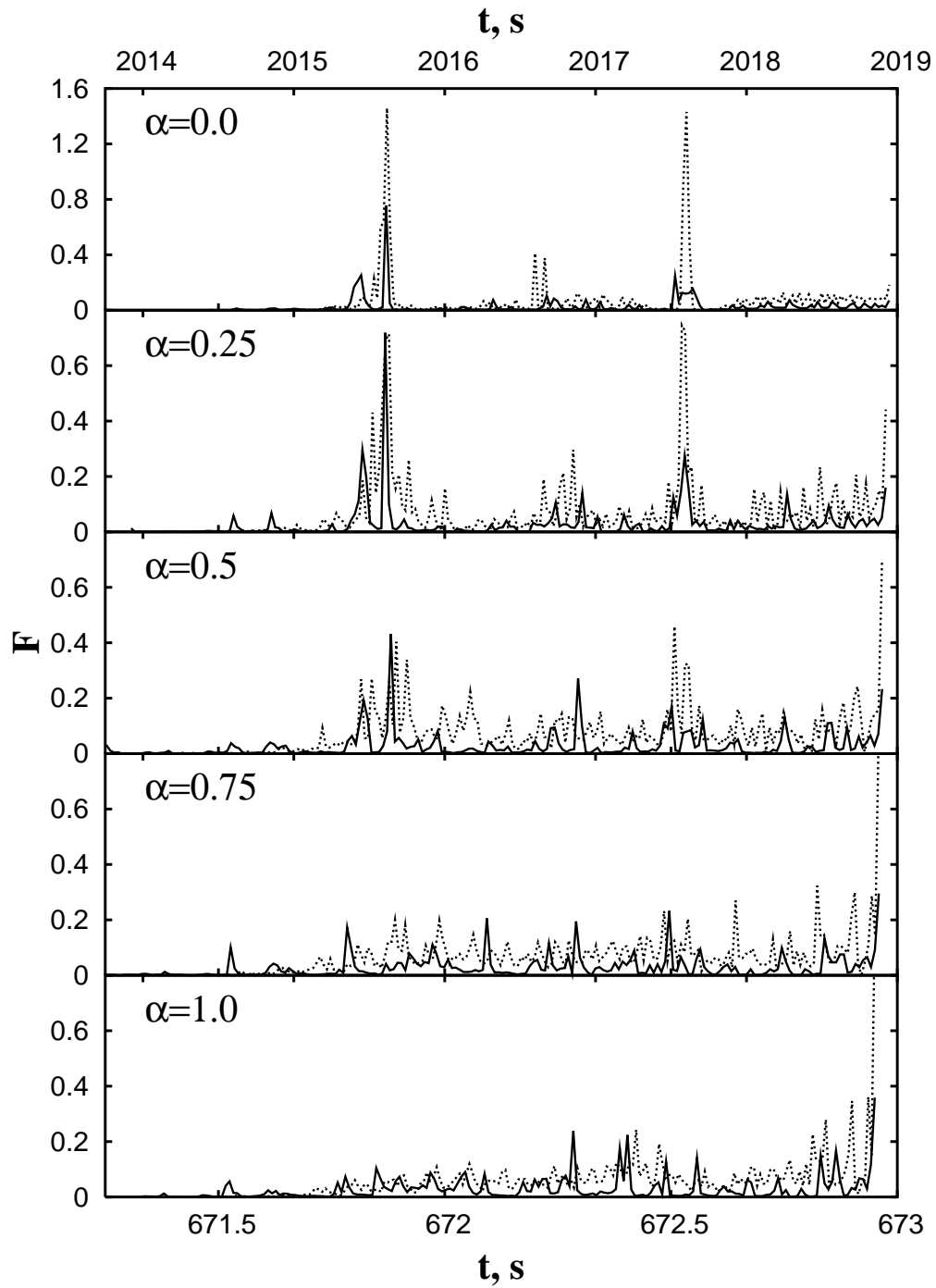}}
\caption{Normalized function of distribution of ray arrival times
for Model~1 with the periodic perturbation ($\lambda=10$~km
and $\varepsilon=0.0025$) and an imposed multiplicative noise with different
values of its strength $\alpha$. The solid lines represent results at
the range 1000 km (see the lower axis for travel times), while the dashed lines
are computed at the range 3000 ~km (see the upper axis for travel times).
The upper panel shows the function with a purely
periodic perturbation.}
\label{fig9}
\end{figure}

The upper panel in Fig.~\ref{fig9} shows the function $F(t,\,R)$
corresponding to the timefront computed at $R=1000$ ~km with Profile 1,
the perturbation wavelength $\lambda=10$~km, the perturbation strength 
$\varepsilon=0.0025$ and the other parameters to be specified
in the preceding section. Results at the range $R=1000$ ~km are 
represented in this figure by solid lines with the lower axis showing 
the respective values of travel times. 
Local concentrations of points in the respective timefront (not shown), 
correspond to two sharp peaks of the function $F(t,\,R)$ 
at $t_\text{res1}\simeq 671.84$~s and $t_\text{res2} 
\simeq 672.54$~s, with the left one corresponding to the resonance
$(l=1,\ m=6)$ and the right
one belonging to the resonance $(l=1,\ m=5)$. Using the formula (\ref{dt}),
one can estimate the respective
resonant cycle lengths, $D(t_\text{res1})\simeq60.6$~km and
$D(t_\text{res2})\simeq50.0$~km, and find numerically the perturbation
wavelength, $\lambda_\text{cal}\simeq 10.6$~km. Note that the upper panel in 
Fig.~\ref{fig9} demonstrates an additional smaller peak at 
$t\simeq 672.25$~s corresponding to the second-order resonance
with $l=2$ and $m=11$ that
is manifested on the Poincar\'e map in Fig.~\ref{fig5} as a chain of islands 
between the first-order resonant islands. It has a comparatively small amplitude
because the width of this high-order resonance and the frequency of phase oscillations
are comparatively small. The satellite peak of the primary
resonance $(l=1,\ m=6)$ seems
to be formed by chaotic rays sticking for a long distance to respective resonant
islands but quitting this zone
somewhere. Note that this peak was absent when we have computed
the distribution function of ray arrival times at $R=3000$~km.

\begin{figure}[!htb]
\centerline{\includegraphics[width=\pictwidth,clip]{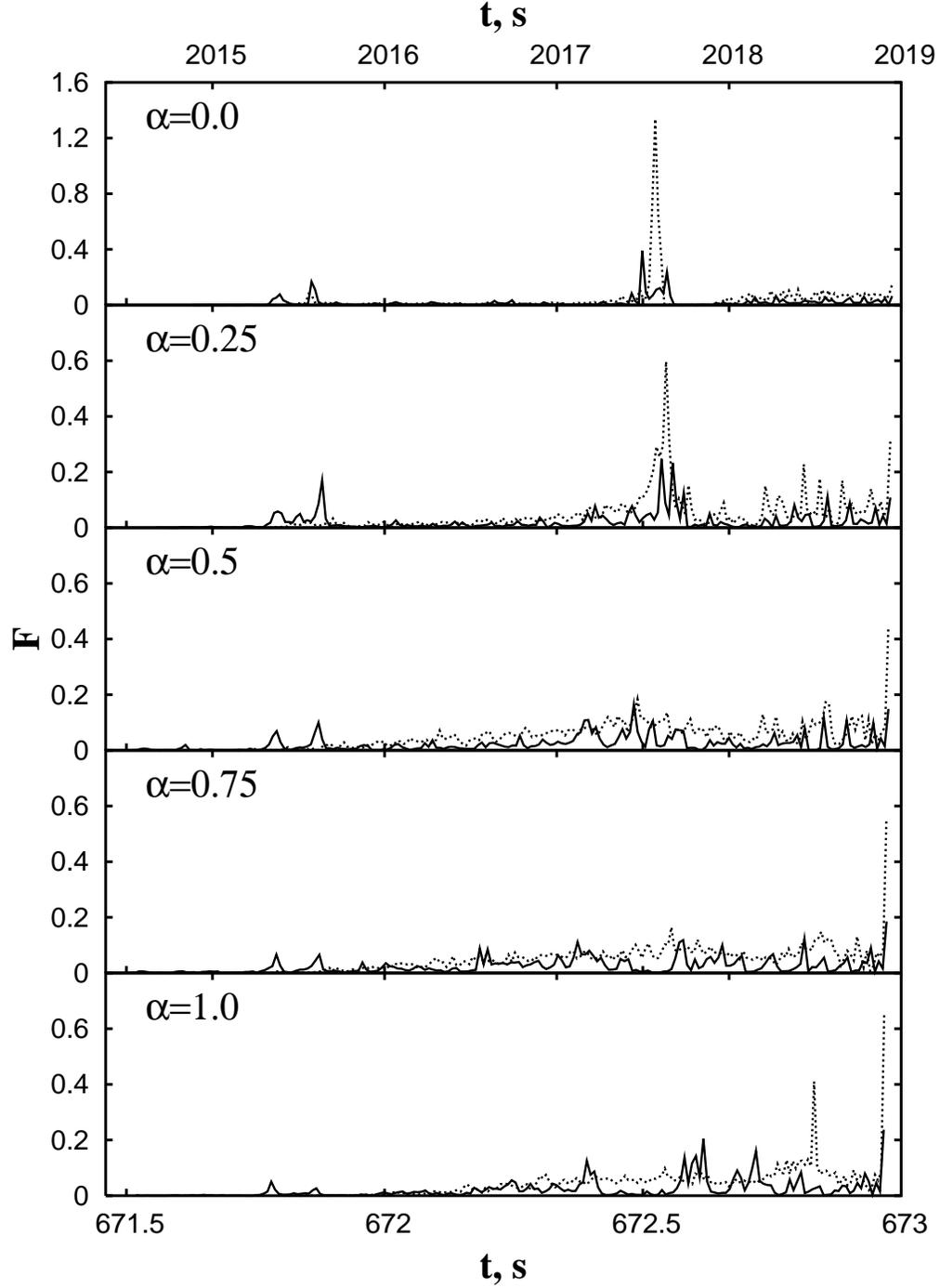}}
\caption{
The same as in Fig.~\ref{fig9} but with the parameters of the periodic
perturbation, $\lambda=10$~km and $\varepsilon=0.005$.
}
\label{fig10}
\end{figure}

\begin{figure}[!htb]
\centerline{\includegraphics[width=\pictwidth,clip]{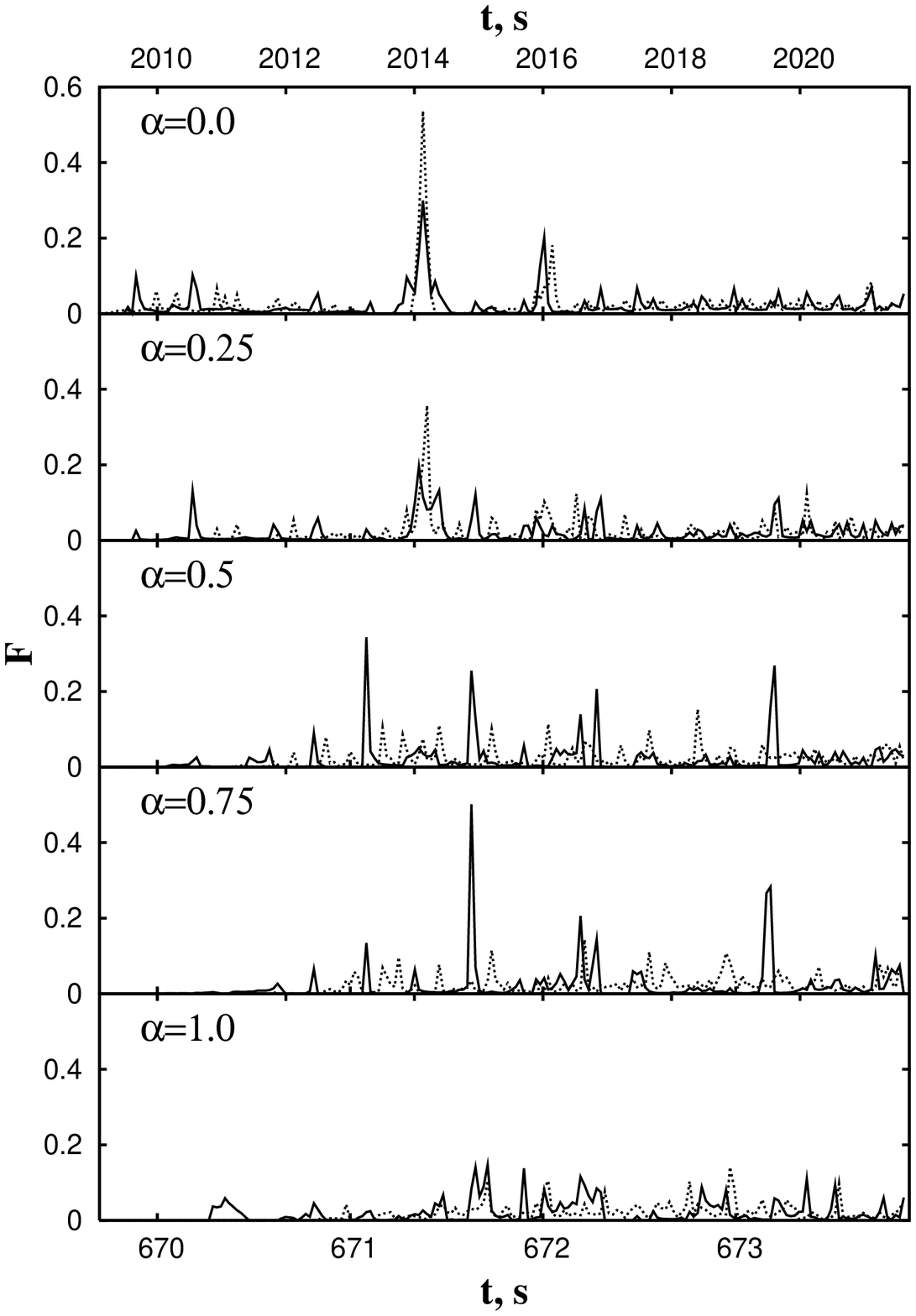}}
\caption{
The same as in Fig.~\ref{fig10} but for Model~2.
}
\label{fig11}
\end{figure}

The upper panel in Fig.~\ref{fig10} shows the function $F(t,\,R)$ with 
Model~1 at the increased value of the perturbation amplitude
$\varepsilon=0.005$ for which ray chaos is stronger.
Because of a large overlapping of the nonlinear resonances, the peak 
corresponding to the resonance with $l=1$ and $m=6$ 
has a smaller amplitude than the respective peak in Fig.~\ref{fig9} 
and disappears at $R=3000$~km at all.
The peak, corresponding to the higher-order resonance with $l=2$ and $m=11$
is absent in Fig.~\ref{fig10}.
The upper panel in Fig.~\ref{fig11} shows the function $F(t,\,R)$ with
Model~2 at $\varepsilon=0.005$. The left peak corresponds to the 
resonance ($l=1$, $m=5$) which is seen on the Poincar\'e map
in Fig.~\ref{fig6} whereas the right peak corresponds to the stochastic layer
near $I(H_r)$. Since diffusion inside this layer is localized the respective 
rays have close arrival times and form a cluster. 

\begin{figure}[!htb]
\centerline{\includegraphics[width=\pictwidth,clip]{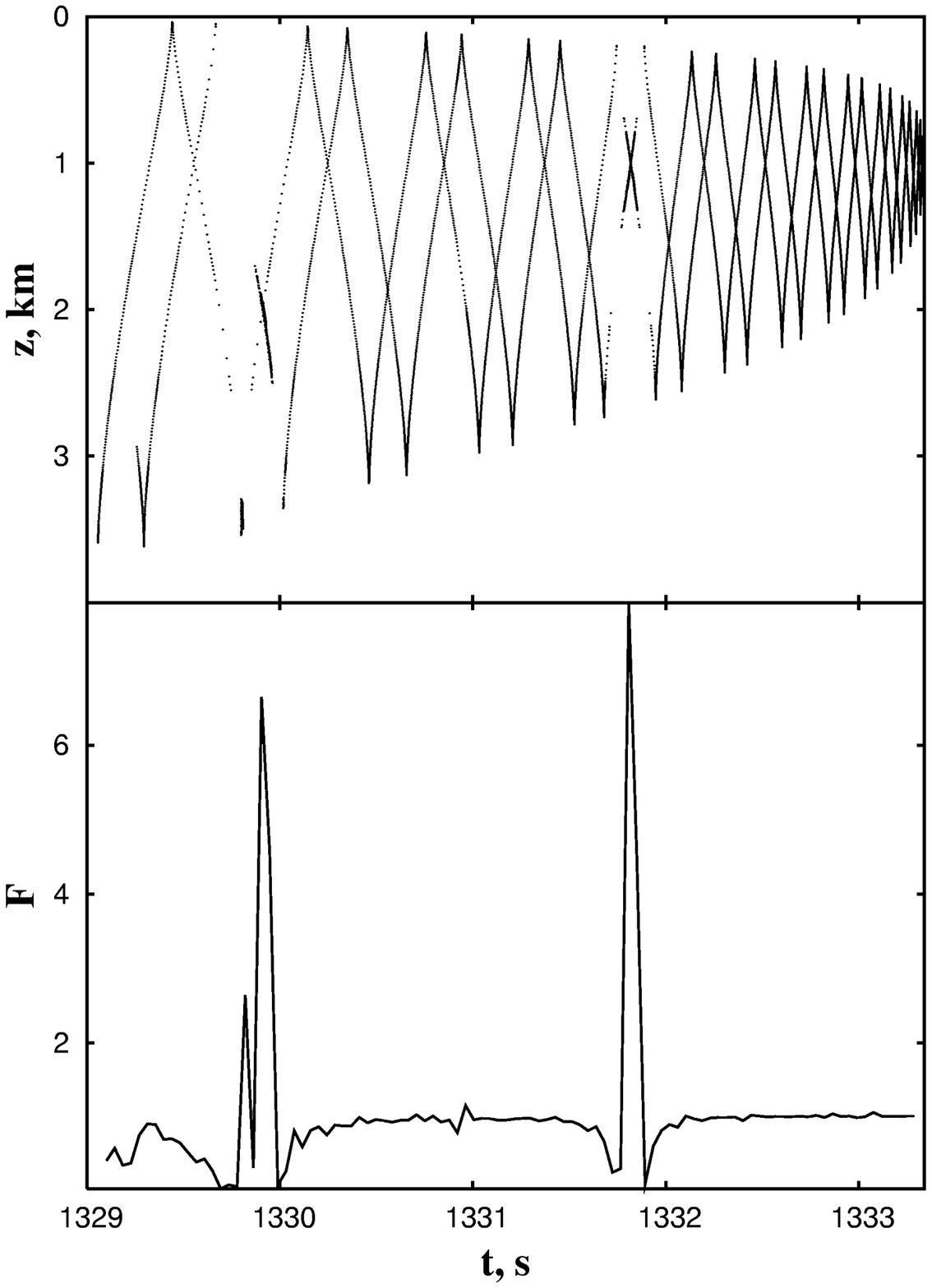}}
\caption{
Timefront and the corresponding function of distribution of ray arrival times
for the Munk canonical profile with the parameters of the periodic
perturbation,
$\lambda=5$~km and $\varepsilon=0.005$.
}
\label{fig12}
\end{figure}

In order to demonstrate the possibility of determining the wavelength of 
an internal wave from the ray arrival time
distribution under conditions of ray chaos not only with our model background
profiles, we have computed the timefront and the respective function
$F(t,\,R)$ with the Munk canonical background profile
(Fig.~\ref{fig2}) with a periodic perturbation \cite{Smith1,Zas-fr,Wierc}
\begin{equation}
c(z,\,r)=c_0\left[1+\mu\,(\eta-1+e^{-\eta})
+\varepsilon\frac{z}{B}\,e^{-2z/B}\cos{\frac{2\pi r}{\lambda}}\right],
\label{Munk}
\end{equation}
where $c_0=1500$~m/s, $\mu=0.0057$, $\eta=2(z-z_a)/B$ is a normalized depth, 
and $B=1$~km. 
The depth of the channel axis is chosen to be $z_a=1$~km, $\varepsilon=0.005$ 
and $\lambda=5$~km.
The results, shown in Fig.~\ref{fig12}
at the range $R=2000$~km, reveal two distinct peaks of the function $F(t,\,R)$,
the left one at $t_\text{res1}\simeq1329.9$~s
corresponds to $D(t_\text{res1})\simeq55$~km and the right one at
$t_\text{res2}\simeq1331.8$~s corresponds to
$D(t_\text{res2})\simeq50$~km giving the difference to be equal
to the perturbation wavelength $\lambda=5$~km.
The resonant cycle lengths with the Munk profile have been found
numerically with the help of Eq.~(\ref{t_un}).
Generally speaking, determining perturbation wavelength is
not always possible even in the case
of a single-mode perturbation. It is necessary to have at least
two chains of slightly overlapped primary
resonances with $l=1$, and sufficiently large number of rays should
be captured in these resonances.

Due to stable motion of rays, captured in a nonlinear resonance,
their initial and final values of the action are in
a comparatively narrow interval $\delta I$ approximately proportional
to the resonance width $\Delta I$.
In accordance with Eq.~(\ref{zti}), depths of arrivals of the
resonant rays at a given range $R$ are distributed
over narrow intervals. In such a way, double sharp strips (see Fig.~\ref{fig12}a),
corresponding to ray clusters with positive and negative
launch angles, appear in the respective timefronts.

\section{\label{secNoise}Ray motion in the presence of a multiplicative noise} 
\subsection{\label{ssecNoiseEq}Ray equations with noise} 
 
Internal waves in the deep ocean are known to have a broadened 
continuous spectrum of horizontal wavenumbers $k$ which may be 
adequately described by the empirical Garrett--Munk spectrum \cite{GM1} 
with the spectral energy density decreasing with increasing $k$. 
The periodic internal-wave induced perturbation of the sound speed 
to be considered in the preceding section is a useful but 
not realistic approximation. From the theoretical point of view, 
the sound-speed perturbations due to internal waves should be 
considered as a random function with given statistical characteristics. 
Under a noisy-like perturbation, there are no specified resonances 
in ray dynamics which, however, is strongly influenced by the presence 
of a nonlinear sound-speed profile \cite{Beron-Vera}. 
 
Let $\xi(r)$ be a stationary stochastic perturbation representing 
the sound-speed fluctuations caused by internal waves which 
is defined as a spectral decomposition 
\begin{equation} 
\xi(r)=\int\limits_{-\infty}^{\infty} S(k)e^{-ikr}dk,
\label{shum} 
\end{equation} 
where $S(k)=S_0(k)\,e^{-i\phi(k)}$ and $\phi(k)$ is a random function 
of the wave number $k$ distributed equally over the interval $[0:2\pi]$. 
The perturbation is assumed to be a Gaussian process with normalized 
first and second moments 
\begin{equation} 
\aver{\xi(r)}=0,\quad 
\aver{\xi^2(r)}=\frac{1}{2}. 
\label{avers} 
\end{equation} 
We assume that the horizontal scale of the internal-wave induced fluctuations 
is much less than the scale of the range variations of the action, and the 
diffusion approximation can be adopted. The perturbation amplitude 
depends on the angle variable $\vartheta$ and is maximal near the upper 
turning point, $\vartheta=0$. So, the ray cycle length $D$ 
gives us a characteristic scale of the range variations of the action. 
In simulation, we realize the process $\xi(r)$ as a sum of a large number 
$(\ge1000)$ of harmonics distributed in the range 
$k\in[2\pi/100:2\pi/1]$~km$^{-1}$. 
Two spectral models of the sound-speed fluctuations, 
$S_0(k)=\text{const}$ and $S_0(k)\propto k^{-2}$, have been used. 
 
The Hamilton equations of motion 
\begin{equation} 
\dfrac{dI}{dr}=-\varepsilon\,\frac{\partial V}{\partial \vartheta}\,\xi(r),\quad 
\dfrac{d\vartheta}{dr}=\omega(I)+\varepsilon\,\frac{\partial V}{\partial I}\,\xi(r) 
\label{dot_can1} 
\end{equation} 
provide the description of sound-ray trajectories through the deep ocean 
with a broad spectrum of internal waves inducing the respective spectrum 
of the sound-speed fluctuations. Substituting the Fourier decomposition 
(\ref{v_ryad}) in the first equation (\ref{dot_can1}) and taking into account 
that $\vartheta=\omega(r)r+\vartheta_0$, we can write down the variation 
of the action over the period as 
\begin{equation} 
\Delta I=\varepsilon 
\sum_{m=1}^\infty mV_{m}(I)\Omega_m, 
\label{int_i} 
\end{equation} 
\begin{equation} 
\Omega_{m}=-\frac{ie^{im\vartheta_0}}{2} 
\int\limits_0^{\infty}\int\limits_0^D 
S(k)e^{i(m\omega-k)r}dk\,dr+\text{c.c.} 
\label{mn} 
\end{equation} 
In order to find $\omega(r)$ let us rewrite the second equation 
(\ref{dot_can1}) as 
\begin{equation} 
\omega(r)-\omega(I_0)=\varepsilon\frac{dV}{dI}\,\xi(r), 
\label{delta_w} 
\end{equation} 
where $I_0$ is the initial value of the action. 
Assuming the correlation length of the random process $\xi(r)$ to be small 
as compared with the ray cycle length, $r_{\xi}\ll D$, 
we introduce the ``fast'' variable $x=r/r_{\xi}$ which is connected with 
the angle variable as follows: 
\begin{equation} 
x=\frac{\vartheta-\vartheta_0}{\omega r_\xi}. 
\label{xtheta} 
\end{equation} 
The variable $\omega(x)$ is now treated as a function of $x$ and modelled 
as a Markovian process with independent increments characterized 
by the normal distribution 
\begin{equation} 
\upsilon\left[w(x),\,I_0\right]=\frac{1}{\sqrt{2\pi}\sigma} 
\exp\left(-\dfrac{\left(\omega-\omega(I_0)\vphantom{\bigl(}\right)^2}{2\sigma^2}\right), 
\label{Gauss} 
\end{equation} 
with the variance depending on $x$ and $\vartheta_0$ 
\begin{equation} 
\sigma^2(x,\,\vartheta_0)=\frac{\varepsilon^2}{2}\int\limits_0^{x} 
\left(\frac{dV(x',\,\vartheta_0)}{dI}\right)^2\,dx'. 
\label{sigma} 
\end{equation} 
The main contribution to the integral (\ref{mn}) provides the points  
of a stationary phase given by the wavenumbers $k_m=m\omega(x)$. So we get  
\begin{equation}
\Omega_m=\pi\aver{S_m[k_m(x)]\cos[m\vartheta_0-\phi[k_m(x)]]}_{k_m(x)}=
\pi\int S_m[k_m(x)]\cos[m\vartheta_0-\phi[k_m(x)]]\rho(k_m)\,dk_m,
\label{omega1}
\end{equation}
where $\rho(k_m)$ is the respective probability density. 
Because the perturbation  $V(\vartheta)$ and its derivative $dV/dI(\vartheta)$ 
have a sharp maximum in a neighbourhood of $\vartheta=0$ and are approximately 
zero outside, we can assume $\rho(k_m)\simeq\upsilon(k_m)$ at  
$\sigma=\sigma(x=-\vartheta_0/\omega r_\xi)$. The integral (\ref{mn}) 
is now given by  
\begin{equation} 
\Omega_{m}=\pi S_{m}^\text{eff}\cos{(m\vartheta_0-\phi_{m}^\text{eff})}, 
\label{f_mn1} 
\end{equation} 
where $\phi_{m}^\text{eff}$ and $S_{m}^\text{eff}$ are an effective phase and 
amplitude, respectively. The amplitude is 
\begin{equation} 
S_{m}^\text{eff}= 
\int\limits_0^\infty 
S_0(k_{m})\,\rho(k_{m})\,dk_{m}, 
\label{S_m} 
\end{equation} 
where $k_{m}=m\omega$. 
As a result, we find the variation of the action over the ray period 
\begin{equation} 
\Delta I=\varepsilon\pi 
\sum_{m=1}^\infty 
mV_{m}(I)S_{m}^\text{eff}\cos{(m\vartheta_0-\phi_{m}^\text{eff})}, 
\label{d_i1} 
\end{equation} 
as a sum of ``resonant'' terms. 
 
\begin{figure}[!htb] 
\centerline{\includegraphics[width=\pictwidth,clip]{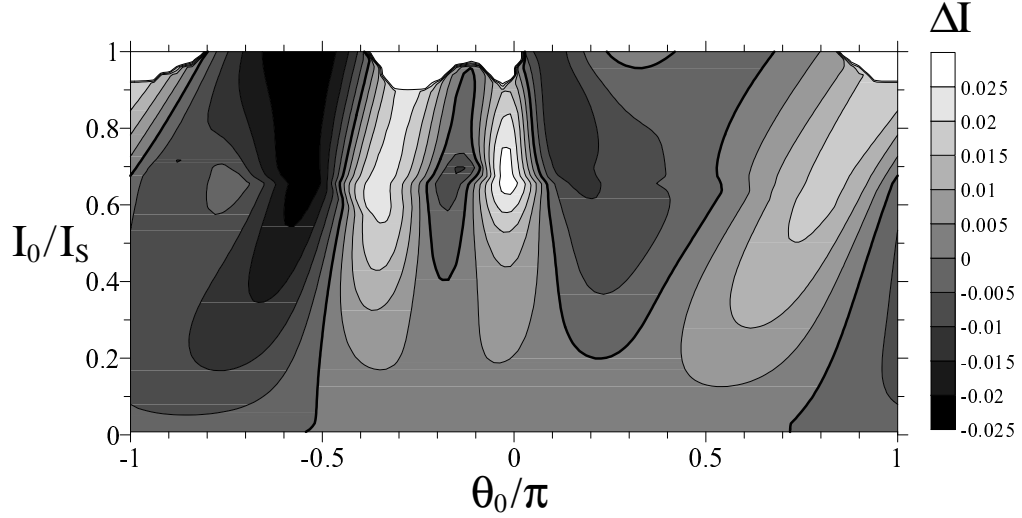}} 
\caption{Plot representing variations of the action $\Delta I$ 
per ray cycle length for Model~2 under the noisy-like perturbation with 
$S_0(k)$ = const and $\varepsilon=0.005$. Bold lines correspond to zero 
variations of the action per ray cycle length. 
} 
\label{fig13} 
\end{figure}

As in the case with a single-mode perturbation, we compute 
plots which show by color modulation 
values of variations of the action per a ray cycle length, $\Delta I$, in the 
plane of the normalized initial values of the action and angle variables. 
The plot in Fig.~\ref{fig13} is computed with Model~2 
and the flat spectrum $S_0(k)=\text{const}$ ($\varepsilon=0.005$). 
To visualize the borders between positive (``hills'') and negative 
(``hollows'') values of $\Delta I$, the lines with $\Delta I=0$ are bolded 
in Fig.~\ref{fig13}. In the range of comparatively small values of the action, 
the first Fourier harmonic $V_1$ in the expansion (\ref{d_i1}) is 
expected to be dominant. Really, only one ``hill'' is present in the figure 
in the range $0<I_0/I_s<0.2$. With increasing the action values, the higher-order 
terms in (\ref{d_i1}) begin to play a more significant role, and the number of ``hills'' 
is expected to rise in the respective ranges on the 
plot representing variations of the action per ray cycle length. 
In Fig.~\ref{fig13} we see two ``hills'' in the range $0.2<I_0/I_s<0.4$ 
and three ``hills'' in the range $0.4<I_0/I_s<0.9$. 
 
\begin{figure}[!htb] 
\centerline{\includegraphics[width=\pictwidth,clip]{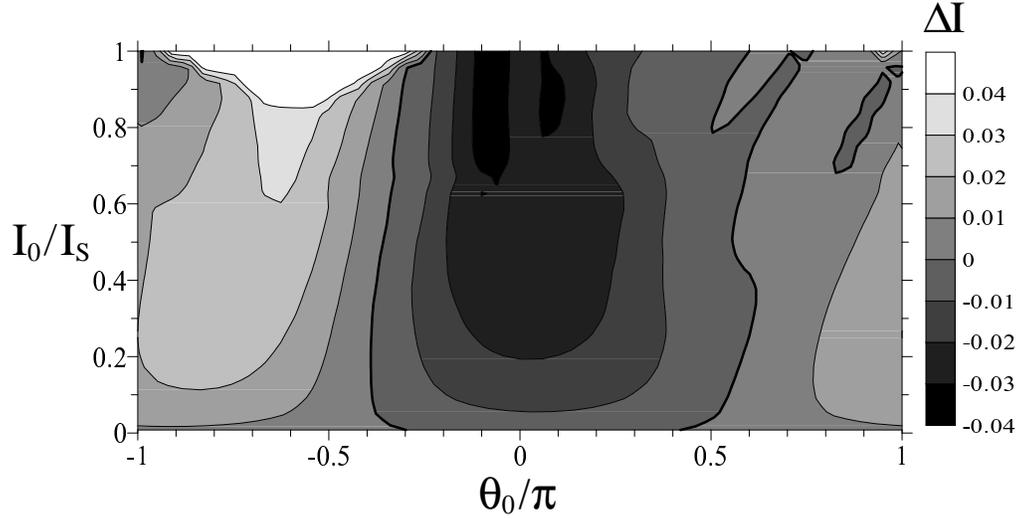}} 
\caption{The same as in Fig.~\ref{fig13} but with 
$S_0(k)\propto k^{-2}$ and $\varepsilon=0.005$. Bold lines correspond to 
zero variations of the action per ray cycle length. 
} 
\label{fig14} 
\end{figure} 
 
Consider now Model~2 with another kind of the perturbation spectrum, 
$S_0\propto k^{-2}$, and the same perturbation amplitude $\varepsilon=0.005$. 
In difference from the case with the flat spectrum, only one large 
``hill'' now presents in Fig.~\ref{fig14} in the whole range of the 
action values. Moreover, the maximal variations of the action are larger 
as compared with the flat spectrum. One may conclude that under the 
noisy-like perturbation with the spectrum $S_0\propto k^{-2}$ the first 
harmonic in the series (\ref{d_i1}) is a dominant one in all the 
accessible phase space. The dependence $\Delta I(\vartheta_0)$ is close 
to a cosine-like one. 
 
Under a noisy-like perturbation, topology of the plots of 
variation of the action depends randomly 
on initial values of the time-like variable $r$, and this dependence is stronger in 
the case of the flat spectrum, especially in the range of large values of 
the action. It may cause not only phase shifts but even the number of 
``hills'' and ``hollows'' may vary under varying initial values of $r$. 
In the case with decreasing spectral density, only smooth shifting of a ``hill'' 
along the $\vartheta$ axis may occur when varying initial values of $r$.

\begin{figure}[!htb] 
\centerline{\includegraphics[width=\pictwidth,clip]{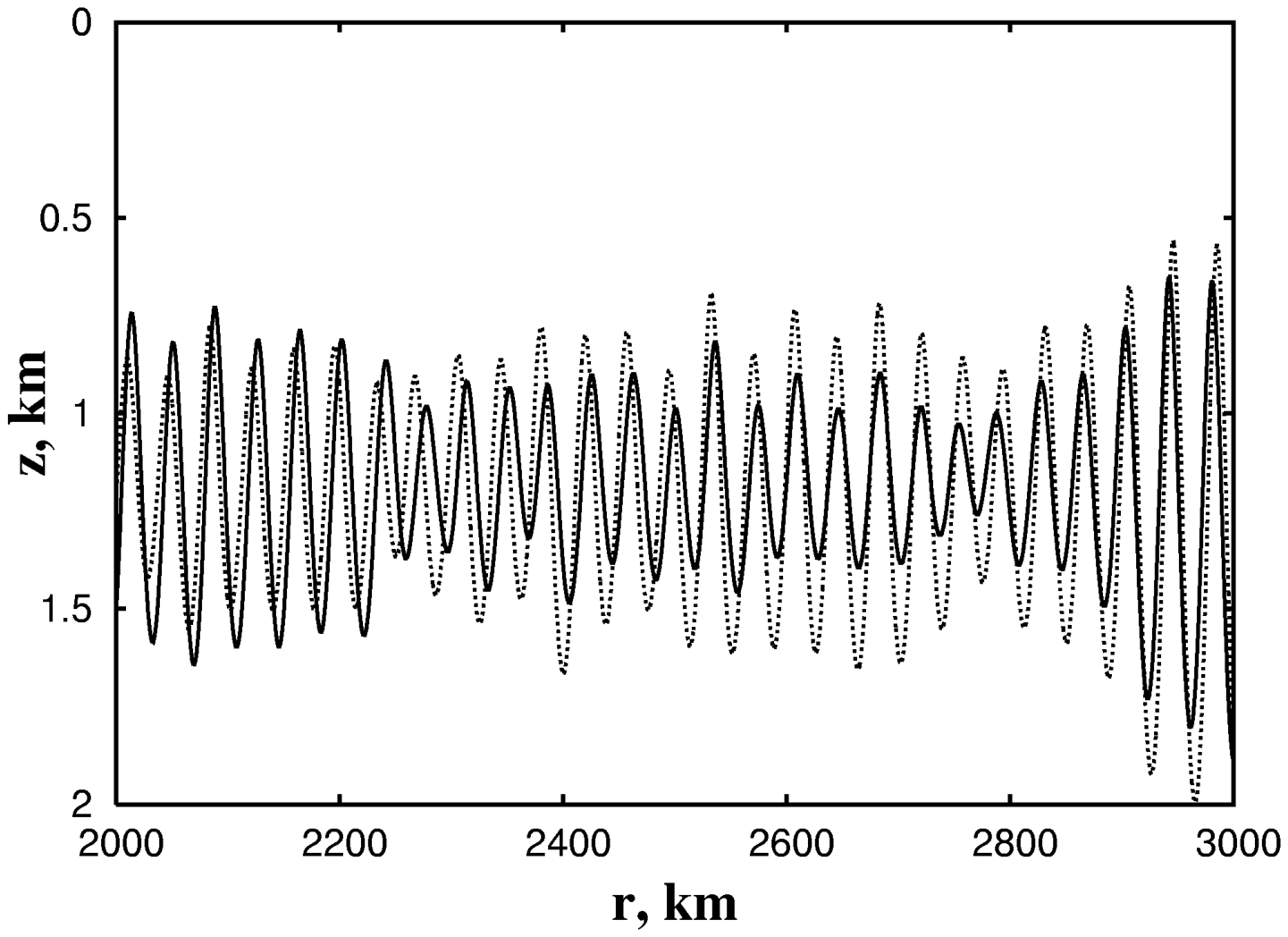}} 
\caption{ 
Two ray trajectories with starting momenta $p_0 = -0.03$ (solid line) 
and $-0.02$ (dashed line) for Model~2 under the noisy-like perturbation 
with $S_0(k)\propto k^{-2}$. 
} 
\label{fig15} 
\end{figure} 
 
\subsection{\label{ssecNoiseClust}Coherent ray clusters}
 
\begin{figure}[!htb] 
\centerline{\includegraphics[width=\pictwidth,clip]{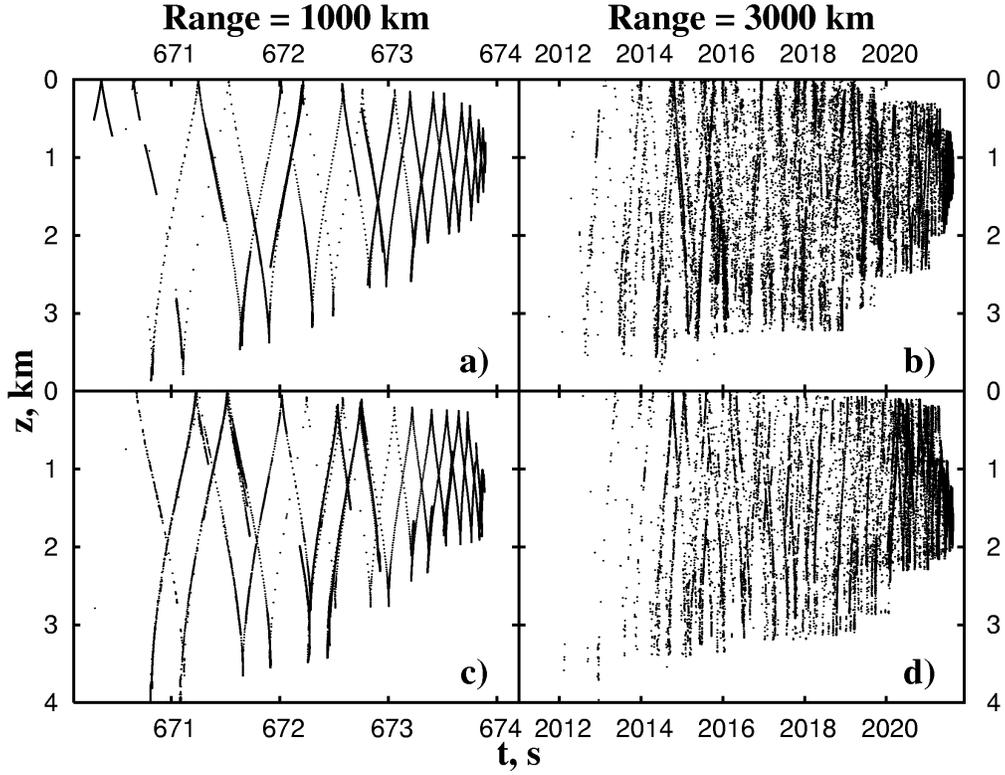}} 
\caption{ 
Timefronts for Model~2 under the noisy-like perturbation with 
(a,b) $S_0(k)$ = const and (c,d) $S_0\propto k^{-2}$. The ranges 
for the left and right plots are indicated in the figure. 
} 
\label{fig16} 
\end{figure}

The plots of variations of the action allow us to treat the ray motion 
as a slow diffusion in the phase space between ``hills'' and ``hollows''. 
If these ``hills'' (or ``hollows'') are sufficiently large there may arise  
large fans of rays with 
close initial conditions preserving close current dynamical characteristics 
over long distances {\it coherent ray clusters}.  In the ranges of strong 
variability of the phase space structure, 
phase correlations decay rapidly resulting in rapidly decaying 
clusters. So, the length of the phase correlations characterizes the 
stability of a cluster. Similar clusterization may occur in different physical 
systems (see, for example, \cite{Klyats,Wolfs2}). Therefore, 
the whole cluster structure may be considered as consisting of statistical and 
coherent parts. The rays, belonging to the statistical part, propagate 
in the same areas of the phase space with the same value of the 
Lagrangian $\aver{L}$, do not correlate with each other and demonstrate 
exponential sensitivity to initial conditions. 
To the contrary, the rays in the coherent part do not show sensitive dependence 
on initial conditions. Two rays with initial values of the momentum 
$p_0=-0.02$ and $p_0=-0.03$ are shown in Fig.~\ref{fig15} in the range interval 
$r\in[2000:3000]$~km. The clusterization may influence 
strongly timefronts of sound signals. The prominent stripes 
visible in the timefront for the stochastic ray simulation (Fig.~\ref{fig16}), 
which belong to ray clusters, resemble the respective strips 
visible in the timefront fragments for a deterministic perturbation 
(see Fig.~\ref{fig12}a). It should be emphasized that similar strips have 
been found in the field experiments \cite{AET}. 
 
A decoherence and breaking of the respective coherent clusters 
become prominent with increasing the range (see fuzzy segments 
in Fig.~\ref{fig16}b at the range $R=3000$~km). 
It is seen from Fig.~\ref{fig16}d that late-arriving 
rays are registered not at the channel axis ($z_a\simeq1$~km) but rather 
deeper, at $z\simeq1.5$~km. Such a shift of the sound energy 
down in the depth may be explained as follows. The late-arriving signal 
is formed by a coherent cluster with near-axial rays deflected under propagation 
from the axial value of the action $I=0$. It follows from Eq.~(\ref{zti}) 
that rays in this coherent cluster could arrive (at $R=3000$~km) at the depth 
different from $z_a$. 
 
\begin{figure}[!htb] 
\centerline{\includegraphics[width=\pictwidth,clip]{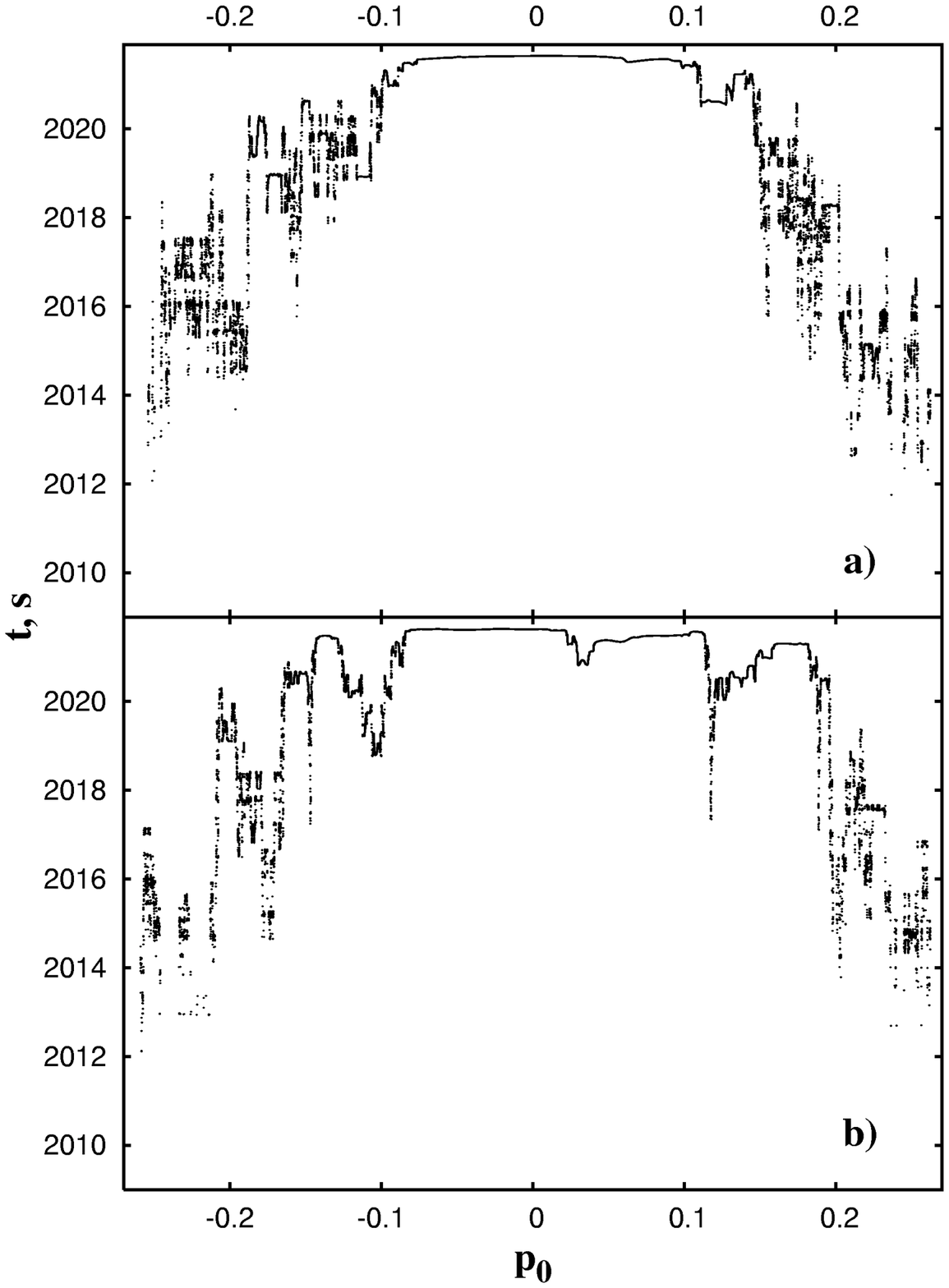}} 
\caption{ 
Ray travel time versus starting momentum for Model~2 
under noisy-like perturbation with (a) $S_0(k)$ = const 
and (b) $S_0\propto k^{-2}$. 
} 
\label{fig17} 
\end{figure} 
 
Figure \ref{fig17} presents the ray travel time $t$ as a function 
of the starting ray momentum $p_0$ for Model~2 under the 
noisy-like perturbation with both the spectral models, 
$S_0(k)=$ const (a) and $S_0(k)\propto k^{-2}$ (b). All rays are chaotic 
under a noise perturbation, and one might naively expect to see randomly 
scattered points in the $t$--$p_0$ plots. In fact, we see in Fig.~\ref{fig17} 
smooth ``shelf''-like segments alternating with unresolvable structures. 
Each ``shelf'' corresponds to a coherent cluster of rays. The ``shelves'' 
are distributed chaotically over the range of the starting momenta 
and their positions depend on 
a specific realization of the random process $\xi(r)$. Comparing between the 
two spectral models (Figs. \ref{fig17}a and b), we may conclude that the 
coherent cluster structure is more prominent with the spectral model $S_0(k)\propto k^{-2}$. 
Such a ``shelf''-like structure has been found in $t(p_0)$ plots for a model 
with a single-mode perturbation \cite{Smir}, with ``shelves'' to be prescribed 
to regular islands in the respective phase space. 
The presence of ``shelves'' may complexify kinetic description of the ray 
motion with the help of a one-dimensional Fokker-Plank equation 
\cite{Viro-web, Zas-PhA} because the 
radius of phase correlations is not small in the presence of coherent 
clusterization. 
 
\subsection{\label{ssecNoisePer}Periodic perturbation with a multiplicative noise superimposed} 
 
In the end of this section we consider a perturbation consisting 
of a periodic dependence of the sound-speed fluctuations 
on $r$ and a multiplicative noise superimposed 
\begin{equation} 
\xi(r)=(1-\alpha)\cos{\frac{2\pi r}{\lambda}}+\alpha\int\limits_{-\infty}^{\infty}
 S(k)e^{-ikr}\,dk, 
\quad
0\le\alpha\le1, 
\label{mix} 
\end{equation} 
where $\alpha$ is the strength of the noisy part. We have computed 
timefronts of sound signals at different values of $\alpha$. 
Fig.~\ref{fig9} shows function $F(t,\,R)$ 
with Model~1 at the fixed ranges $R=1000$~km and $R=3000$~km 
with the perturbation amplitude $\varepsilon=0.0025$ and 
spatial period $\lambda=10$~km corresponding to the Poincar\'e section 
in Fig.~\ref{fig5}. The solid lines in Figs.~\ref{fig9}--\ref{fig11}
represent results at 
the range 1000 km with the lower axis showing the respective values of 
travel times, while the dashed lines 
are computed at the range 3000 ~km with the upper axis for travel times. 
As it expected, the amplitudes of the prominent peaks, caused by the 
nonlinear resonance with the periodic perturbation, 
decreases with increasing the values of $\alpha$. 
On the other hand, the amplitudes of the peaks in the late-arriving signal, 
caused by the noisy-like perturbation, increases with increasing 
$\alpha$. 
 
Fig.~\ref{fig10} shows the function $F(t,\,R)$ for Model~1 under conditions of 
more strong chaos at increased value of the perturbation amplitude 
$\varepsilon=0.005$. 
All the ``deterministic'' peaks have comparatively 
small amplitudes even at $\alpha=0.5$. 
The distribution function $F(t,\,R)$ is shown for Model~2 in
Fig.~\ref{fig11} with the same values of the parameters
of perturbation as 
in the preceding figure with Model~1. It is seen that the peaks at $t\simeq371.3$~s 
($R=1000$~km) and at $t=2014$~s ($R=3000$~km), corresponding to the primary 
resonance of the first order $(l=1,\ m=5)$, disappear when the strength of 
noise reaches the magnitudes of the order $\alpha \ge 0.5$. 
 
In difference from Model~1, coherent ray clusters in Model~2 appear not only in 
the late-arriving portion of the signal but, as well, in the early arriving 
portion corresponding to rays reflecting from the ocean surface. 
The rays with $H>H_r$ in Model~2 are less chaotic than the rays with 
$H < H_r$. These results show that ray dynamics is strongly influenced by 
the form of the background sound-speed profile. 
Depending on the form of the background profile, coherent ray clusters may appear 
in earlier, middle and later portions of a timefront. In our opinion, 
a stability in earlier portions of the wavefront to be measured in the 
field experiments \cite{Duda, Worc3250} could be explained by peculiarities 
of the respective background sound-speed profile.

Under a deterministic perturbation with only a few frequencies, chaoticity 
of rays is defined 
mainly by the density of overlap of nonlinear resonances characterized by 
the Chirikov's criterion (\ref{Chi}) and is connected with the derivative of 
the frequency of spatial oscillations over the action 
$|\omega'_I|$ (\ref{dwi}). However, the Chirikov's criterion is hardly 
applicable under conditions of a noisy-like perturbation with a large 
number of frequencies \cite{Beron-Vera, Geoph}. In this case we propose to use 
as a criterion of stochasticity the rate 
of decreasing of Fourier amplitudes in the series (\ref{d_i1}). 
Stochasticity of rays 
has been shown to become stronger if many terms present in the 
series (\ref{d_i1}). It can be used as a 
stochasticity criterion for nonlinear systems under a noisy-like 
perturbation in a close analogy with the Chirikov's criterion for deterministic 
dynamical systems since the rate of decreasing of Fourier amplitudes is 
defined mainly 
by the dependence of the frequency of spatial oscillations $\omega$ on 
the action $I$. It is a linear function with Model~1 (\ref{wI}). 
In Model~2 the respective dependence $\omega(I)$ has a local maximum 
at $I_r\simeq I(H_r)$. Thus, there arise conditions for forming 
coherent ray clusters in Model~2.
 
\section{\label{SecConc}Conclusion} 
 
We have treated chaotic and stochastic nonlinear ray dynamics 
in underwater sound waveguides 
with longitudinal variations of the speed of sound caused by internal 
oceanic waves. Two models of sound-speed profiles, which are typical in 
shape for deep ocean sound channels, were designed analytically. 
We were managed to 
derive with them exact analytical solutions to the ray equations of motion 
without perturbation in terms of the depth-momentum and the action-angle
variables and to find exact expressions for the frequency of 
spatial ray oscillations, the timefront of the sound signal at a fixed range 
and ray travel times. Three different kinds of internal-wave induced 
perturbations have been considered: a single-mode perturbation, a noisy-like 
multiplicative perturbation, and a periodic perturbation with a multiplicative 
noise superimposed. 
 
We have found  coherent 
clusters consisting of fans of rays with close dynamical characteristics over 
long distances and close arrival times. It is 
essential that forming the coherent clusters occurs under 
different kinds of perturbations, as periodic as noisy-like ones. 
The mechanism of their forming 
has been found to be connected with existence of specific zones of 
stability in the phase space of the perturbed system under consideration. 
In the case of a periodic perturbation, these zones appear due to ray-medium
nonlinear resonances. In the case of a noisy-like multiplicative 
perturbation, zones of stability appear due to selective resonant interactions 
between different spectral components of the perturbation and harmonics of 
the unperturbed motion. As a result, the phase space has a specific 
``resonant'' topology with local zones of stability. In order to visualize 
the topology, we have used the plots of variations of the action per ray cycle length. 
 
We proposed a criterion for forming 
the coherent clusters, namely, the rate of decreasing of the Fourier 
amplitudes of 
a perturbation written in terms of the canonical action and angle 
variables. The effect of coherent clusterization depends on the horizontal spectrum of 
the field of internal waves as well. The clusterization becomes more prominent 
if the spectral density decreases rapidly with increasing the wave number $k$. 
The clusterization results in forming prominent peaks of functions of distribution 
of arrival times and manifests itself in timefronts of arriving 
signals as sharp strips on a smearing background formed by chaotic rays. 
It is worthwhile to stress that such strips have been found in timefronts measured 
in the field experiments \cite{AET}. The clusterization may cause a redistribution 
of the acoustic energy over the depth, a stability of early arriving part of 
the sound signal and other effects. It should be taken into account in kinetic 
modelling of ray dynamics. 
 
From the standpoint of acoustic tomography of the ocean, the coherent clusterization 
seems to be a useful property for the purpose of determining spatio-temporal variations of 
the hydrological characteristics on the real time scale under conditions 
of ray chaos. In a more general context, such a clusterization is interesting 
from the standpoint of general theory of influence of external multiplicative
noise on Hamiltonian systems. 
 
\section*{Acknowledgments} 
 
This work was supported by the Program ``Mathematical Methods in 
Nonlinear Dynamics'' of the Russian Academy of Sciences, by the Russian 
Foundation for Basic Research (03--02--06896), and by the Program for 
Basic Research of the Far Eastern Division of the Russian Academy of 
Sciences.

\end{document}